\documentclass[conference]{IEEEtran}
\usepackage[T1]{fontenc}
\usepackage[latin1]{inputenc}
\usepackage{color}
\usepackage{psfrag}
\usepackage[dvips]{graphicx}
\usepackage{tikz}
\usepackage{pgfplots}
\pgfplotsset{compat=1.12}
\usepackage{enumerate}
\usepackage{rotating}
\usepackage{srcltx}
\usepackage{booktabs}
\usepackage{enumitem}
\usepackage[noadjust]{cite}
\usepackage[cmex10]{amsmath}
\usepackage{amsfonts}
\usepackage{amssymb}
\usepackage{ntheorem}
\usepackage{dsfont}    %mathds offers better set-symbols for N,Z,Q,R,C
\usepackage{mathtools} %\coloneqq for perfect :=
\usepackage{stackengine}

\usepackage{algorithmic}
\usepackage{array}
\usepackage{graphicx}
\usepackage{caption}
\usepackage{subcaption}

\theoremseparator{.}
\newtheorem{corollary}{Corollary}
\newtheorem{prop}{Proposition}
\newtheorem{lem}{Lemma}
\newtheorem{theorem}{Theorem}
\theorembodyfont{\upshape}
\theoremseparator{.}
\newtheorem{definition}{Definition}
\newtheorem{remark}{Remark}
\newtheorem*{example*}{Example}

\newcommand{\eu}{\mathrm{e}}

\newcommand{\mmse}{\mathrm{mmse}}

\newcommand{\E}{\mathbb{E}}

\newcommand{\supp}{{\mathsf{supp}}}

\long\def\symbolfootnote[#1]#2{\begingroup%
	\def\thefootnote{\fnsymbol{footnote}}\footnote[#1]{#2}\endgroup} %unnumbered footnote for thanks
\title{On the Capacity of the Peak Power Constrained Vector Gaussian Channel:
An Estimation Theoretic Perspective} 
\author{\IEEEauthorblockN{Alex~Dytso\IEEEauthorrefmark{1}, Mert~Al\IEEEauthorrefmark{2},  H. Vincent~Poor\IEEEauthorrefmark{3} and Shlomo Shamai (Shitz)\IEEEauthorrefmark{4}\vspace{5pt}}%
\IEEEauthorblockA{\IEEEauthorrefmark{1},\IEEEauthorrefmark{2},\IEEEauthorrefmark{3}Department of Electrical Engineering, Princeton University\\
				  \IEEEauthorrefmark{4}Department of Electrical Engineering, Technion -- Israel Institute of Technology}  %
				   \IEEEauthorblockA{ Email: adytso@princeton.edu\IEEEauthorrefmark{1}, merta@princeton.edu\IEEEauthorrefmark{2}, poor@princeton.edu\IEEEauthorrefmark{3} and sshlomo@ee.technion.ac.il\IEEEauthorrefmark{4}}}%%
\begin{document}
\maketitle
\symbolfootnote[0]{This work was supported in part by the U. S. National Science Foundation under Grant CNS-1702808,  and in part by the 
European Union's Horizon 2020 Research And Innovation Programme, grant agreement no. 694630.}
%
%%%%%%%%%%%%%%%%%%%%%%%%%%%%%%%%%%%%%%%%%%%%%%%%%%%%%%%%%%%%%%%%%%%%%%%%%%%%%%%%%%%%%%%%%%%%%%%%%%%%%%%%%%%%%%%%%%%%%%%%%%%
%%%%%%%%%%%%%%%%%%%%%%%%%%%%%%%%%%%%%%%%%%%%%%%%%%%%%%%%%%%%%%%%%%%%%%%%%%%%%%%%%%%%%%%%%%%%%%%%%%%%%%%%%%%%%%%%%%%%%%%%%%%
%	Abstract
%%%%%%%%%%%%%%%%%%%%%%%%%%%%%%%%%%%%%%%%%%%%%%%%%%%%%%%%%%%%%%%%%%%%%%%%%%%%%%%%%%%%%%%%%%%%%%%%%%%%%%%%%%%%%%%%%%%%%%%%%%%
%%%%%%%%%%%%%%%%%%%%%%%%%%%%%%%%%%%%%%%%%%%%%%%%%%%%%%%%%%%%%%%%%%%%%%%%%%%%%%%%%%%%%%%%%%%%%%%%%%%%%%%%%%%%%%%%%%%%%%%%%%%
\begin{abstract}%
This paper studies the capacity of an $n$-dimensional vector Gaussian noise channel  subject to the constraint that an input must lie in the ball of radius $R$ centered at the origin.  It is known that in this setting the optimizing input distribution is supported on a finite number of concentric spheres. However, the number, the positions and the probabilities of the spheres are  generally unknown. This paper characterizes   necessary and sufficient conditions on the constraint $R$   such that the input distribution  supported on a single sphere is optimal. The maximum $\bar{R}_n$,  such that using only a single sphere is optimal, is shown to be  a solution of an integral equation.   Moreover, it is shown that  $\bar{R}_n$   scales as $\sqrt{n}$ and the exact limit of $\frac{\bar{R}_n}{\sqrt{n}}$ is found.  
\end{abstract}%

\section{Introduction}
We consider  an additive noise channel for which the input-output relationships are given by 
\begin{align}
Y=X+Z, \label{eq:channel}
\end{align}
where $X \in \mathbb{R}^n$ is independent of $Z \in  \mathbb{R}^n$ and where $Z \sim \mathcal{N}(0,I_n)$.   We are interested in finding the capacity of the channel  in \eqref{eq:channel} subject to the constraint that $X \in  \mathcal{B}_0(R) $ where  $\mathcal{B}_0(R)$ is  an  $n$-ball  centered at $0$ of radius $R$ (amplitude or peak power constraint), that is
\begin{align}
\max_{X  \in \mathcal{B}_0(R)} I(X;Y). \label{eq:Capacity}
\end{align}
In general the capacity in \eqref{eq:Capacity} is an open problem and only some special cases have been solved.  In this work the capacity in \eqref{eq:Capacity} will be characterized  for all $R$ that are smaller than roughly  $\sqrt{n}$. 

  The necessity of characterization of the capacity with a peak power constraint on the input is self-evident. Many practical systems inherently have a peak power constraint  due to the limited range of operations of electronic equipment. Some channels  (e.g.,  the direct detection photon channel \cite{shamai1990capacity}) have well defined ranges of operations where average power constraints are not relevant and peak power constraints must be used.

\subsection{Prior Work}

For the case of $n=1$ Smith in his seminal work \cite{smith1971information}, using convex optimization techniques, has shown that the maximizing distribution in \eqref{eq:Capacity} must be discrete with finitely many points. In \cite{ShamQuadrat}, for the case of $n=2$,  the maximizing input distribution has been shown to be supported on  finitely many concentric spheres.   The generalization to  an arbitrary $n$   can be found in \cite{chan2005MIMObounded,rassouli2016capacity} and \cite{CISS2018}.

This paper can be considered as an $n$-dimensional generalization of the work in \cite{sharma2010transition} where, in the case of $n=1$ and under the conjecture that the number of mass points, as we vary $R$, increases by at most one,  a two point input distribution  uniform  on $\pm R$ has been shown to be optimal if and only if $R \le 1.665$,  and a three point input distribution on $\{-R,0,R \}$ has been shown to be optimal if and only if $ 1.665 \le R \le  2.786$. However, unlike the approach in \cite{sharma2010transition}, the proof strategy used in this work relies on very different  methods (rooted in estimation theory) and, for every dimension $n$, recovers the exact condition for the optimality of an input supported on a single sphere.  Moreover, our proof does not require the assumption of the conjecture that the number of points increases by at most one as we vary $R$. 

 The fact that a uniform distribution on a single sphere is optimal as $ \frac{R}{\sqrt{n}} \to 0$ has been shown in \cite{rassouli2016capacity}. Moreover, the authors of \cite{rassouli2016capacity} have  observed via numerical results the fact that a distribution  with the support on a single sphere can be optimal for  non-vanishing values of $R$. In addition,  the authors of  \cite{rassouli2016capacity}  have  computed   the maximum values of  $R$, for which a single sphere is optimal,  up to $n=20$.  

A number of works have also focused on deriving lower and upper bounds on \eqref{eq:Capacity}. The authors in \cite{thangaraj2017capacity} derived an asymptotically tight  upper bound on the capacity  as $R\to \infty$  by using the dual representation of channel capacity.  In \cite{DytsoGlobcome} the authors derived an upper bound on the capacity, by using a maximum entropy principle under $L_p$ moment constraints, that is tight for small values of $R$.
See also  \cite{DytsoGlobcome}  and \cite{elmoslimany2016capacity} for asymptotically tight lower bounds on the capacity. 

The interested reader is also referred to \cite{polyanskiy2014peak} where in addition to the amplitude input constraint the authors also considered an average power constraint on the input and characterized the amplitude-to-power ratio of good codes.

\subsection{Paper Outline and  Contributions}
The paper outline and contributions are as follows:
\begin{enumerate}
\item    Section~\ref{sec:OptimalInputDistribution} reviews some known facts about the optimal input distribution in \eqref{eq:Capacity} (e.g., the support is given by concentric spheres);
\item   Section~\ref{sec:DefinitionOfAsmallAmplitudeRegime}  gives the definition of the ``small amplitude" regime  as the regime in which a uniform probability distribution supported on a single sphere is optimal;
\item  Section~\ref{sec:MainResult}, Theorem~\ref{thm:MainResult}, presents our main result, which is an exact characterization of the size of the small amplitude regime.  The proof of the main result is postponed to Section~\ref{sec:ProofOfTheMainResult};
\item  Section~\ref{sec:SomeAnalyticComputations}, for an input distribution on $X$ uniformly distributed on a sphere of radius $R$,  computes the output distribution, the conditional expectation of the  input  $X$ given the output $Y$, the  mutual information between  $X$ and $Y$ and the minimum mean square error (MMSE) of estimating $X$ from $Y$;
\item  Section~\ref{sec:NewCondition} presents  new conditions for the optimality of the distribution on a single sphere. The new conditions have an advantage of   being  easier to verify than the classical conditions presented in Section~\ref{sec:OptimalInputDistribution}. The key ingredients for the proof  of the new conditions  are the change of sign lemma due to Karlin \cite{karlin1957polya},  the I-MMSE relationship \cite{I-MMSE} and the point-wise I-MMSE relationship \cite{pointWiseI-MMSE}.  To the best of our knowledge this is the first application of the  point-wise I-MMSE relationship to a capacity problem;
\item Section~\ref{sec:ProofOfTheMainResult}  presents the proof of the main result;
\item Section~\ref{sec:AlternativeProof} gives an alternative proof, using yet another information estimation identity,  that $R \le \sqrt{n}$ is sufficient for the optimality of the distribution on a single sphere; and 
\item Section~\ref{sec:Discussion}  concludes the paper by discussing connections  between  maximization of the mutual information and 
 maximization of the MMSE (i.e., the theory of  finding least favorable prior distributions). In particular, we discuss conditions under which least favorable distributions are also capacity achieving. 
 \end{enumerate} 

\subsection{Definitions and Notation}

The volume of the unit $n$-ball and the unit $(n-1)$-sphere are denoted and given by 
\begin{align}
V_n&\coloneqq \frac{\pi^{\frac{n}{2}}}{\Gamma \left(\frac{n}{2}+1 \right)},\\
S_{n-1}&\coloneqq {\frac {2\pi ^{\frac {n}{2}}}{\Gamma \left({\frac {n}{2}}\right)}}.
\end{align}
 We denote the $(n-1)$-sphere of radius $r$ centered at the origin  as follows:
\begin{align}
\mathcal{C}(r)\coloneqq \{ x : \|x \|=r \}. 
\end{align}

  $Q(\cdot)$  denotes the tail distribution function of the standard normal distribution. The modified Bessel function of the first kind of order $v$ is  denoted by  $ \mathsf{I}_{v}(x)$.  We also use the following commonly encountered ratio of Bessel functions:
\begin{align}
\mathsf{h}_v(x)\coloneqq \frac{ \mathsf{I}_{v}(x)}{ \mathsf{I}_{v-1}(x)}. 
\end{align}

We denote the distribution  of a random variable $X$ by $P_X$. Moreover, we say that a point $x$ is in the support of the distribution $P_X$ if for every open set $\mathcal{O}$ such that $x \in \mathcal{O}$ we have that $P_X(\mathcal{O})>0$ and denote the collection of the support points of $P_X$ as $\supp(P_X)$. 

At times it will be convenient to use the following parametrization of the mutual information in terms of the input distribution  $P_X$: 
\begin{align}
 I(P_X)\coloneqq I(X;Y).
\end{align}
We also define the following quantity that is akin to the information density: 
\begin{align}
i(x, P_X) &\coloneqq \int_{\mathbb{R}^n}  \frac{1}{ (2 \pi)^{\frac{n}{2}}}  \eu^{-\frac{\| y-x\|^2}{2}} \log \frac{1}{f_Y(y)} {\rm d}y -h(Z)\\
&= \E \left[ \log \left( \frac{f_{Y|X}(Y|X)}{f_Y(Y)}  \right)  \mid X=x \right],
\end{align}
where  $f_Y(y)$ is the output probability density function (pdf) of $Y$ induced by $X \sim P_X$ and $h(Z)$ is the entropy of Gaussian noise.  Moreover, note that
\begin{align}
\E[i(X, P_X)]= I(P_X). 
\end{align}

The MMSE of estimating the input $X$ from the  output $Y$ will be denoted as follows:
\begin{align}
\mmse(X \mid Y)\coloneqq\E \left[\| X -\E[X \mid Y]\|^2 \right].
\end{align}

\section{Optimizing The Input Distribution}
\label{sec:OptimalInputDistribution}

The optimal input distribution in \eqref{eq:Capacity} can be characterized by using the method presented in \cite{smith1971information} and its extension to the complex channel (i.e., $n = 2$)  given in \cite{ShamQuadrat}; see also \cite{chan2005MIMObounded,rassouli2016capacity} and \cite{CISS2018} %this is us
 for a detailed solution for any $n \ge 2$.  

\begin{theorem} \emph{(Characterization of the Optimal Input Distribution)} \label{thm:OptimalityCondition} Suppose $P_X^\star$ is an optimizer in \eqref{eq:Capacity}. Then,   $P_X^\star$ satisfies the following properties: 
\begin{itemize}
\item $P_X^\star$ is unique;
\item  $P_X^\star$ is optimal if and only if  the following two conditions are satisfied: 
\begin{subequations}
\begin{align}
i(x, P_X^\star) &=  I(P_X^\star),   \,   x  \in \supp(P_X^\star), \label{eq:equalityCondition}\\
i(x, P_X^\star)  &\le   I(P_X^\star), \,  x \in  \mathcal{B}_0(R); \text{ and} \label{eq:inequalitityCondition}
\end{align} 
\item the support of the optimal input distribution is given by 
\begin{align}
\supp(P_X^\star)= \bigcup_{i=1}^N  \mathcal{C}(r_i), \label{eq:Support}
\end{align}
where $N<\infty$ (finite).
\label{eq:SuffAndNecc}
\end{subequations}
\end{itemize}
\end{theorem}

\begin{figure}
\center
\input{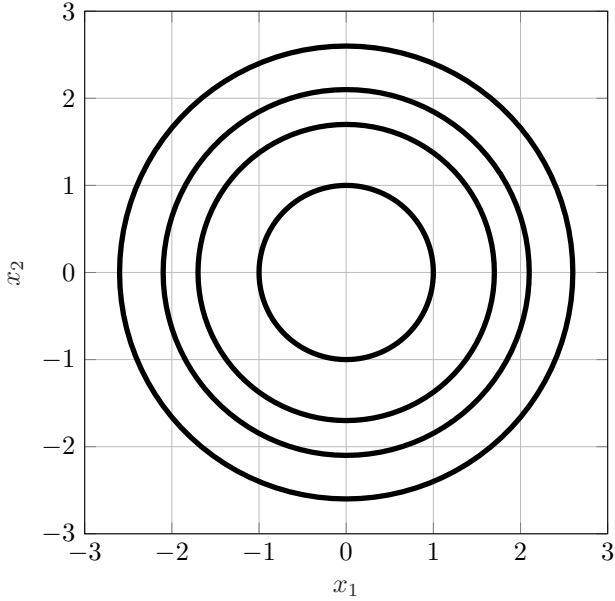}
\caption{An example of a support of an optimal input distribution for $n=2$.}
\label{fig:InputSphere}
\end{figure}

An example of the support of  distributions in \eqref{eq:Support} for $n=2$ is  shown in Fig.~\ref{fig:InputSphere}.

Note that for $n=1$ the optimal inputs are discrete with finitely many points. For $n>1$ the optimal input probability distributions are no-longer discrete but singular, however, the magnitude of the optimal input distribution $\|X\|$ is discrete with finitely many points.

\subsection{Small Amplitude Regime}
\label{sec:DefinitionOfAsmallAmplitudeRegime} 
In this paper the small amplitude regime has the following definition.  

\begin{definition*} Let  $X_R \sim P_{X_R}$  be uniform on $\mathcal{C}(R)$. 
 The capacity in \eqref{eq:Capacity} is said to be in \emph{the small amplitude regime}  if   $R \le  \bar{R}_n$  where
\begin{align}
 \bar{R}_n \coloneqq  \max \{ R :   P_{X_R}=\arg\max \max_{X  \in \mathcal{B}_0(R)} I(X;Y)     \}.
\end{align}
In words,  $ \bar{R}_n $ is the largest radius $R$ for which $P_X$ uniformly distributed on $\mathcal{C}(R)$ is the  capacity achieving distribution in \eqref{eq:Capacity}.    
\end{definition*}

In this work we are interested in  exactly characterizing $ \bar{R}_n$.  

\section{Main Result} 
\label{sec:MainResult}

 The following theorem, which is the main result of this paper,  gives a complete characterization of the small amplitude regime.

\begin{theorem}  \emph{(Characterization of the Small Amplitude Regime)} \label{thm:MainResult} The input $X_R$ is optimal in \eqref{eq:Capacity} (i.e., capacity achieving) if and only if $R \le \bar{R}_n$ where   $\bar{R}_n$ is given as the solution of the following equation:
\begin{subequations}
\begin{align}
 \int_{0}^1  \hspace{-0.15cm}  \E \left[    \mathsf{h}_{\frac{n}{2}}^2\left(  \sqrt{\gamma} R   \|  Z \| \right) \right] + \E \left[    \mathsf{h}_{\frac{n}{2}}^2\left(  \sqrt{\gamma} R   \|  \sqrt{\gamma} x+ Z \| \right) \right]  {\rm d} \gamma=1, \label{eq:ExactR}
\end{align} 
for any $x$  such that $\|x\|=R$. 
In addition, it is sufficient to take $R \le \sqrt{n}$  (i.e., $\sqrt{n} \le \bar{R}_n$), and 
\begin{align}
 \lim_{n \to \infty} \frac{\bar{R}_n}{\sqrt{n}} = c \approx 1.860935682,
\end{align}
where $c$ is the solution of the following equation:
\begin{align}
\int_0^1  \hspace{-0.16cm} \frac{ \gamma  c^2 }{ \left( \frac{1}{2 }  +\sqrt{ \frac{1}{4 }  + \gamma c^2} \right)^2} \hspace{-0.05cm}  + \hspace{-0.05cm}\frac{  \gamma c^2(1+ \gamma c^2) }{ \left( \frac{1}{2 }  +\sqrt{ \frac{1}{4 }  + \gamma c^2(1+ \gamma c^2)} \right)^2} {\rm d} \gamma =1. \label{eq:SolutionInfinity}
\end{align}
\end{subequations}
\end{theorem}
\begin{IEEEproof}
See Section~\ref{sec:ProofOfTheMainResult}.
\end{IEEEproof}

\begin{figure}

\begin{subfigure}[a]{0.45\textwidth}
\center
\pgfplotsset{every axis plot/.append style={very thick}}
\definecolor{mycolor1}{rgb}{0.00000,0.44700,0.74100}%
\definecolor{mycolor2}{rgb}{0.85000,0.32500,0.09800}%
\begin{tikzpicture}

\begin{axis}[%
width=6.953cm,
height=6.226cm,
at={(1.392in,0.874in)},
scale only axis,
xmin=0,
xmax=35,
xlabel style={font=\color{white!15!black}},
xlabel={$n$},
ymin=1,
ymax=11.7,
axis background/.style={fill=white},
xmajorgrids,
ymajorgrids,
legend style={at={(0.19,0.78)}, anchor=south west, legend cell align=left, align=left, draw=white!15!black}
]
\addplot [color=black,dashed]
  table[row sep=crcr]{%
1	1.66592407226562\\
2	2.45351033957971\\
3	3.06515589156213\\
4	3.57955932617188\\
5	4.03127544570278\\
6	4.4380182276662\\
7	4.81120929712101\\
8	5.15755503369971\\
9	5.48258972167969\\
10	5.78919352468905\\
11	6.08048721693125\\
12	6.35894904411483\\
13	6.62504092126062\\
14	6.88141012571975\\
15	7.12831402147235\\
16	7.36721801757812\\
17	7.59848429248675\\
18	7.82298377368965\\
19	8.04114294320148\\
20	8.25365527284442\\
21	8.460696451575\\
22	8.66280484559857\\
23	8.86027971288904\\
24	9.05361100623514\\
25	9.24282073974609\\
26	9.42835441814296\\
27	9.61017829552024\\
28	9.78870658329041\\
29	9.96402575577891\\
30	10.1364527169396\\
31	10.305707888045\\
32	10.4721633862921\\
33	10.6362852926261\\
34	10.797839926672\\
35	10.9569253203337\\
};\addlegendentry{$\bar{R}_n$}

\addplot [color=black]
  table[row sep=crcr]{%
1	1\\
2	1.4142135623731\\
3	1.73205080756888\\
4	2\\
5	2.23606797749979\\
6	2.44948974278318\\
7	2.64575131106459\\
8	2.82842712474619\\
9	3\\
10	3.16227766016838\\
11	3.3166247903554\\
12	3.46410161513775\\
13	3.60555127546399\\
14	3.74165738677394\\
15	3.87298334620742\\
16	4\\
17	4.12310562561766\\
18	4.24264068711928\\
19	4.35889894354067\\
20	4.47213595499958\\
21	4.58257569495584\\
22	4.69041575982343\\
23	4.79583152331272\\
24	4.89897948556636\\
25	5\\
26	5.09901951359278\\
27	5.19615242270663\\
28	5.29150262212918\\
29	5.3851648071345\\
30	5.47722557505166\\
31	5.56776436283002\\
32	5.65685424949238\\
33	5.74456264653803\\
34	5.8309518948453\\
35	5.91607978309962\\
}; \addlegendentry{$\sqrt{n}$}
\end{axis}
\end{tikzpicture}%
\caption{Comparison of $\bar{R}_n$ and $\sqrt{n}$. }
        \end{subfigure}%
        \vspace{-0.5cm}
        
\begin{subfigure}[c]{0.45\textwidth}
\center
\pgfplotsset{every axis plot/.append style={very thick}}
\begin{tikzpicture}

\begin{axis}[%
width=6.953cm,
height=6.226cm,
at={(1.159in,0.77in)},
scale only axis,
xmin=0,
xmax=35,
xlabel style={font=\color{white!15!black}},
xlabel={$n$},
ymin=1.66,
ymax=1.86,
axis background/.style={fill=white},
xmajorgrids,
ymajorgrids,
legend style={at={(0.69,0.805)}, anchor=south west, legend cell align=left, align=left, draw=white!15!black}
]
\addplot [color=black, dashed]
  table[row sep=crcr]{%
1	1.66592407226562\\
2	1.73489379882812\\
3	1.76966857910156\\
4	1.78977966308594\\
5	1.80284118652344\\
6	1.81181335449219\\
7	1.81846618652344\\
8	1.82347106933594\\
9	1.82752990722656\\
10	1.83070373535156\\
11	1.83333587646484\\
12	1.83567047119141\\
13	1.83745574951172\\
14	1.83913421630859\\
15	1.84052276611328\\
16	1.84180450439453\\
17	1.84290313720703\\
18	1.84389495849609\\
19	1.84476470947266\\
20	1.84557342529297\\
21	1.84627532958984\\
22	1.84691619873047\\
23	1.84749603271484\\
24	1.84806060791016\\
25	1.84856414794922\\
26	1.84905242919922\\
27	1.84947967529297\\
28	1.84989166259766\\
29	1.85027313232422\\
30	1.85065460205078\\
31	1.85095977783203\\
32	1.85123443603516\\
33	1.85153961181641\\
34	1.85181427001953\\
35	1.85205841064453\\
};
\addlegendentry{$\frac{ \bar{R}_n}{\sqrt{n}}$}
\end{axis}
\end{tikzpicture}%
            \caption{Plot of $\bar{R}_n$ normalized by  $\sqrt{n}$. }
        \end{subfigure}%
        
        \caption{Plots of $\bar{R}_n$ as defined in Theorem~\ref{thm:MainResult} vs. $n$.}
\label{fig:ExactRplot}
\end{figure}
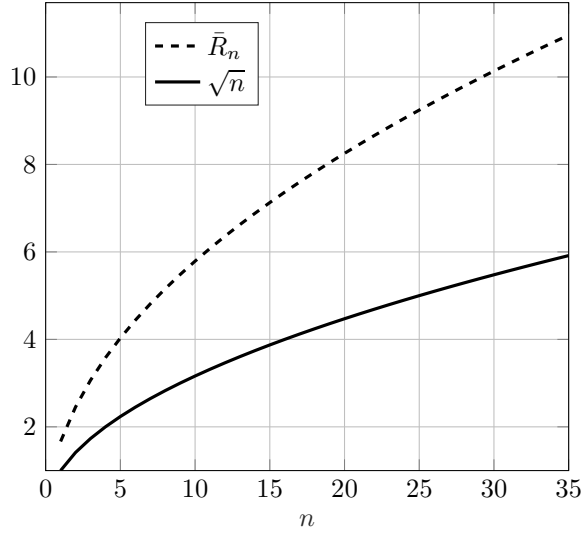
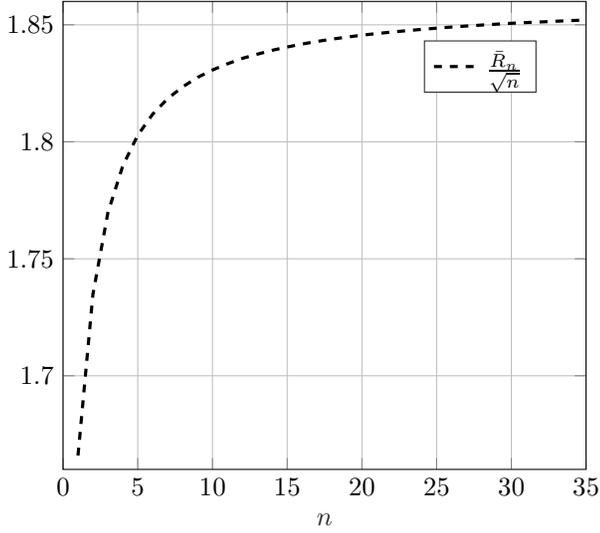

Note that $R_n$ is given as the solution of an integral equation in \eqref{eq:ExactR} and does not have an exact analytical form and must be   found  using numerical methods.  Similarly,  while the integral in \eqref{eq:SolutionInfinity} does have a closed form expression given \eqref{eq:ExpressionForTheIntegration}, the resulting equation must be solved numerically.     The numerical evaluation of $\bar{R}_n$  up to $n=35$ is shown on Fig.~\ref{fig:ExactRplot} and the values of $\bar{R}_n$  are provided in Table~\ref{table:ValuesOfRn}.

It is important to note that numerical computation of $\mathsf{h}_{\frac{n}{2}} \left( x\right)$ via direct evaluations of the Bessel functions may be unstable for large $x$.  The interested reader is referred to Appendix~\ref{app:sec:ComputationOfRatioOfBesselFunctions} for a discussion of these stability issues and details on how values of $\bar{R}_n$    can be computed  by  using a known continued fraction expansion  of $\mathsf{h}_{\frac{n}{2}} \left( x\right)$.

%\begin{table}
%\renewcommand{\arraystretch}{1.3}
%\caption{Values of $\bar{R}_n$ computed within an error of $10^{-4}$.}
%\label{table:ValuesOfRn}
%\centering
%\begin{tabular}{ |c|c| } 
%\hline
%Dimension $n$ &  Maximum Radius $\bar{R}_n$  \\
%\hline
%1  &  1.6659 \\
%\hline
%2  &  2.4529 \\
%\hline
%3  &  3.0652 \\
%\hline
%4  &  3.5797 \\
%\hline
%5  &  4.0316 \\
%\hline
%6  &  4.4372 \\
%\hline
%7  &  4.8108 \\
%\hline
%8  &  5.1574 \\
%\hline
%9  &  5.4829 \\
%\hline
%10  &  5.7893 \\
%\hline
%11  &  6.0805 \\
%\hline
%12  &  6.3581 \\
%\hline
%13  &  6.6249 \\
%\hline
%14  &  6.8812 \\
%\hline
%15  &  7.1285 \\
%\hline
%16  &  7.3677 \\
%\hline
%17  &  7.5984 \\
%\hline
%18  &  7.8227 \\
%\hline
%19  &  8.0409 \\
%\hline
%20  &  8.2533 \\
%\hline
%21  &  8.4606 \\
%\hline
%22  &  8.6628 \\
%\hline
%23  &  8.8602 \\
%\hline
%24  &  9.0542 \\
%\hline
%25  &  9.2431 \\
%\hline
%26  &  9.4283 \\
%\hline
%27  &  9.6103 \\
%\hline
%28  &  9.7887 \\
%\hline
%29  &  9.9640 \\
%\hline
%31  &  10.1361 \\
%\hline
%32  &  10.3052 \\
%\hline
%33  &  10.4726 \\
%\hline
%33  &  10.6360 \\
%\hline
%34  &  10.7979 \\
%\hline
%35  &  10.9572 \\
%\hline
%\end{tabular}
%\end{table}

%\begin{figure}
%\center
%\includegraphics[width=7cm]{FIG/SpheresIn2}
%\input{FIG/RmaxItemp.tex}
%\caption{Plot of $\bar{R}_n$ vs. $\sqrt{n}$. }
%\label{fig:ExactRplot}
%\end{figure}

\begin{remark} Recall that $\| Z+x\|^2$ in \eqref{eq:ExactR} is distributed according to the   non-central chi-square  distribution of degree $n$ with  non-centrality parameter $\| x\|^2$; this fact becomes useful when numerically computing $\bar{R}_n$.   % this is useful for numerical results  
\end{remark}

We can also give the following alternative characterization of $\bar{R}_n$  that does not require  integration over $\gamma$ as in \eqref{eq:ExactR}.

\begin{theorem} \emph{(Alternative Characterization of $\bar{R}_n$)}  \label{thm:AnotherAltertnativeCondtion}  The input $X_R$ is optimal in \eqref{eq:Capacity} if and only if $R \le \bar{R}_n$ where   $\bar{R}_n$ is given as a positive zero of the following equation:  
\begin{align}
\E \left[  \frac{W_1}{ \|W\| } \mathsf{h}_{\frac{n}{2}} \left(  R \|W\|  \right)   \right]  =\frac{1}{2}, \label{eq:SecondCondition}
\end{align}
where $W$ is a random vector of independent components such that $W_1 \sim \frac{  Q(w-R)-Q(w) }{R}$ and $W_i \sim \mathcal{N}(0,1)$  for $2 \le i \le n$. 
\end{theorem}
\begin{IEEEproof}
See Section~\ref{sec:proof:thm:AnotherAltertnativeCondtion}. 
\end{IEEEproof}

\begin{remark} For the case of $n=1$  using the fact that 
\begin{align}
R  \frac{y}{ |y| } \mathsf{h}_{\frac{1}{2}} \left(  R |y|  \right) =\E[X_R|Y=y]= R \tanh(Ry), 
\end{align}
  the expression in \eqref{eq:SecondCondition}  simplifies to  
\begin{align}
\int_{\mathbb{R}}    \left( Q(w-R)-Q(w)  \right) \tanh(  R w)  {\rm d}w =  \frac{R}{2}.  \label{eq:n=1ajflajfa}
\end{align}
The non-zero solution to \eqref{eq:n=1ajflajfa} can be easily found numerically and  is given by $\bar{R}_1 \approx 1.665925641$ as was already computed in \cite{sharma2010transition}.  However, interestingly, while the expression \eqref{eq:n=1ajflajfa} is equivalent to the one presented in  \cite{sharma2010transition},  it is not of the same form. In  \cite{sharma2010transition} $\bar{R}_1$ is  instead given as a solution of the following equation: 
\begin{align*}
& \int_{\mathbb{R}}  \left(  \frac{\eu^{-\frac{ (y-R)^2}{2}}+\eu^{-\frac{ (y+R)^2}{2}}}{2}-\eu^{-\frac{y^2}{2}} \right)  \log \left( \eu^{-Ry}+\eu^{Ry} \right) {\rm d}y \\
 &= \frac{ \sqrt{2 \pi} R^2}{2}. 
\end{align*}
\end{remark}

\section{Some Analytical Computations}
\label{sec:SomeAnalyticComputations}
In this section for the input $X_R$ we compute the output pdf, the mutual information and the MMSE.

\begin{prop} \emph{(Output Distribution)} The pdf of the output distribution induced by the input $X_R$ is given by 
\begin{align}
f_{Y}(y)= \frac{\Gamma \left( \frac{n}{2}\right)   \eu^{ - \frac{R^2+\|y\|^2}{2}} }{ 2 \pi^{\frac{n}{2}}}    \frac{   \mathsf{I}_{ \frac{n}{2}-1}(\|y\| R)}{ (\| y\| R)^{\frac{n}{2}-1}}.  \label{eq:OuputDistribuiton}
\end{align}
\end{prop}
\begin{IEEEproof}
Let $\hat{X}_\epsilon$ have distribution with the pdf given on the annulus
\begin{align}
f_{\hat{X}_\epsilon}(x)=    \frac{1}{V_n  \left( R^n -(R -\epsilon)^n\right)}   1_{ \{ R -\epsilon \le \| \hat{X}_\epsilon \| \le R\} }(x),
\end{align}
for some $\epsilon>0$.
Observe that $\hat{X}_\epsilon \to X_R$ in distribution as $\epsilon \to 0$  and, therefore, by the Dominated Convergence Theorem the output pdf can be written as follows: 
\begin{align}
f_{Y}(y)&= \lim_{\epsilon \to 0}   \E \left[ \frac{1}{ (2 \pi)^{\frac{n}{2}}}  \eu^{-\frac{\| y- \hat{X}_\epsilon \|^2}{2}}  \right] \notag\\
&=    \frac{1}{ V_n (2 \pi)^{\frac{n}{2}}}   \lim_{\epsilon \to 0}   \frac{\int_{ R -\epsilon \le  \|x \| \le R}  \eu^{-\frac{\| y- x \|^2}{2}}  {\rm d}x}{ \left( R^n -(R -\epsilon)^n\right)}.
 \label{eq:limitNeedsToBeComputed}\end{align} 

To compute the limit in \eqref{eq:limitNeedsToBeComputed}  we will need the following integral \cite{watson1983statistics}: % see also \text{https://dlmf.nist.gov/10.32#E2}:
\begin{align}
 \int_{\|x\|=1}\eu^{x^T y R}\mathrm{d}x = \left(\frac{\|y\| R}{2} \right)^{1- \frac{n}{2}}S_{n-1}\Gamma \left(\frac{n}{2} \right)\mathsf{I}_{\frac{n}{2}-1}(\|y\| R). \label{eq:InetegralBasselFunction}
\end{align}
The derivation of the limit in \eqref{eq:limitNeedsToBeComputed} now proceeds as follows:
\begin{align}
&\lim_{\epsilon \to 0} \frac{\int_{ R -\epsilon \le  \|x \| \le R}  \eu^{-\frac{\| y- x \|^2}{2}}  {\rm d}x}{ \left( R^n -(R -\epsilon)^n\right)}  \notag\\
 & \stackrel{a)}{=}  \lim_{\epsilon \to 0} \frac{  \int _{R-\epsilon}^R \int_{ S_{n-1}}  \eu^{-\frac{\| y-r \Theta \|^2}{2}}  r^{n-1} {\rm d}\Theta  {\rm d}r}{ \left( R^n -(R -\epsilon)^n\right)}  \notag  \\
& \stackrel{b)}{=} \lim_{\epsilon \to 0} \frac{  \int _{R-\epsilon}^R f(r,y)  {\rm d}r}{ \left( R^n -(R -\epsilon)^n\right)}  \notag  \\
&\stackrel{c)}{=}\frac{  \frac{d}{d \epsilon} \int _{R-\epsilon}^R f(r,y)  {\rm d}r  |_{\epsilon=0}}{   \frac{d}{d \epsilon}  \left( R^n -(R -\epsilon)^n\right)  |_{\epsilon=0}}  \notag  \\
&\stackrel{d)}{=} \frac{ f(R,y)  }{  nR^{n-1}}  \notag  \\
& = \frac{\int_{ S_{n-1}}  \eu^{-\frac{\|y- R \Theta \|^2}{2}}  R^{n-1} {\rm d}\Theta }{  nR^{n-1}} \notag  \\
&= \frac{   \eu^{- \frac{R^2+\|y\|^2}{2} }  \int_{ S_{n-1}}  \eu^{ R \Theta^T y }   {\rm d}\Theta }{  n }  \notag  \\
&\stackrel{e)}{=}      \frac{  \eu^{- \frac{R^2+\|y\|^2}{2} }  \left(\frac{\|y\| R}{2} \right)^{1- \frac{n}{2}}S_{n-1}\Gamma \left(\frac{n}{2} \right)\mathsf{I}_{\frac{n}{2}-1}(\|y\| R)}{  n } ,\label{eq:FinalLimit}
\end{align}
where the labeled equalities follow from: a) changing to spherical coordinates; b) defining $f(r)\coloneqq \int_{ S_{n-1}}  \eu^{-\frac{\| y-r \Theta \|^2}{2}}  r^{n-1} {\rm d}\Theta$; c) applying  L'H\^opital's rule; d) applying the  Fundamental Theorem of Calculus; and e) using the integral in \eqref{eq:InetegralBasselFunction}.

Putting together \eqref{eq:limitNeedsToBeComputed} and \eqref{eq:FinalLimit} and using \eqref{eq:InetegralBasselFunction} we have that the output pdf is given by 
\begin{align*}
&f_{Y}(y)\\
&=        \frac{1}{V_n  (2 \pi)^{\frac{n}{2}}}   \frac{  \eu^{- \frac{R^2+\|y\|^2}{2} }  \left(\frac{\|y\| R}{2} \right)^{1- \frac{n}{2}}S_{n-1}\Gamma \left(\frac{n}{2} \right)\mathsf{I}_{\frac{n}{2}-1}(\|y\| R)}{  n }\\
& =     \frac{\Gamma \left( \frac{n}{2}\right)   \eu^{ - \frac{R^2+\|y\|^2}{2}} }{ 2 \pi^{\frac{n}{2}}}    \frac{   \mathsf{I}_{ \frac{n}{2}-1}(\|y\| R)}{ (\| y\| R)^{\frac{n}{2}-1}}. 
\end{align*}
This concludes the proof. 
\end{IEEEproof}

\begin{figure}
\input{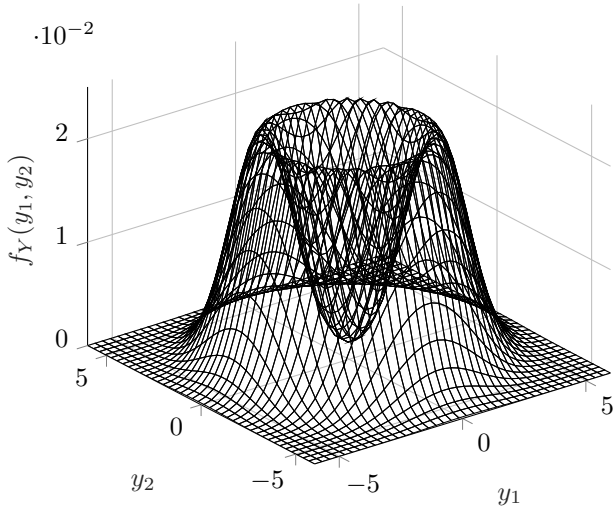}
\caption{The output pdf in \eqref{eq:OuputDistribuiton} for $n=2$ and $R=3$.}
\label{fig:OuputSphere}
\end{figure}

For $n=1$ using the identity $\mathsf{I}_{-\frac{1}{2}}(x)= \left( \frac{2}{\pi x} \right)^{\frac{1}{2}} \cosh(x)	$ we have that 
\begin{align*}
f_{Y}(y)&=  \frac{  \eu^{ - \frac{R^2+|y|^2}{2}} }{ \sqrt{2 \pi} }       \cosh(|y| R)\\
&= \frac{1}{2}  \left(  \frac{1}{ \sqrt{2 \pi} } \eu^{ - \frac{ (R+|y|)^2}{2}} +  \frac{1}{ \sqrt{2 \pi} }  \eu^{ - \frac{ (R-|y|)^2}{2}} \right).
\end{align*}
For $n=2$ the output distribution  is shown in Fig.~\ref{fig:OuputSphere} and  is given by
\begin{align*}
f_{Y}(y)= \frac{  \eu^{ - \frac{R^2+\|y\|^2}{2}} }{ 2 \pi^2}      \int_0^\pi \eu^{\|y\| R \cos(\theta)} {\rm d} \theta;
\end{align*}
for $n=3$ using the identity $\mathsf{I}_{\frac{1}{2}}(x)= \left( \frac{2}{\pi x} \right)^{\frac{1}{2}} \sinh(x)	$ we have that 
\begin{align*}
f_{Y}(y)=  \frac{\sqrt{2}}{8 \pi^{\frac{3}{2}}}  \frac{1}{\|y\| R} \left(   \eu^{ - \frac{ (R-\|y\|)^2}{2}}  -  \eu^{ - \frac{ (R+\|y\|)^2}{2}} \right). 
\end{align*}

Using the expression for the pdf in \eqref{eq:OuputDistribuiton} we can now also compute the conditional expectation  $\E[X|Y]$.

\begin{prop} \emph{(Conditional Expectation)} \label{prop:ConditionExpectation}   For every $R>0$
\begin{align}
\E[X_R \mid Y=y]=   \frac{R y}{\|y\|}    \mathsf{h}_{\frac{n}{2}}\left(  \|y\| R \right).\label{eq:ConditionalExpectation}
\end{align} 
\end{prop}
\begin{IEEEproof}
Using the identity between the conditional expectation and score function \cite{esposito1968relation} we have that 
\begin{align}
\E[X_R \mid Y=y]=y + \frac{ \nabla_y f_Y(y)}{f_Y(y)},  \label{eq:ScoreFunction}
\end{align}
and due to the symmetry of  $f_Y(y)$ we have that
\begin{align}
\nabla_y f_Y(y)&= \frac{y}{\|y\|}  \frac{d}{d\|y\|} f_Y(\|y\|) ,  \label{eq:ChainRule}
\end{align}
  where
\begin{align}
 &\frac{d}{d\|y\|} f_Y(\|y\|) \notag\\
 &=  \frac{d}{d\|y\|}  \frac{\Gamma \left( \frac{n}{2}\right)   \eu^{ - \frac{R^2+\|y\|^2}{2}} }{ 2 \pi^{\frac{n}{2}}}    \frac{   \mathsf{I}_{ \frac{n}{2}-1}(\|y\| R)}{ (\| y\| R)^{\frac{n}{2}-1}} \notag\\
&= - \frac{\Gamma \left( \frac{n}{2}\right)   \eu^{ - \frac{R^2+\|y\|^2}{2}}   \| y \|}{ 2 \pi^{\frac{n}{2}}}    \frac{   \mathsf{I}_{ \frac{n}{2}-1}(\|y\| R)}{ (\| y\| R)^{\frac{n}{2}-1}}  \notag\\
&+  \frac{\Gamma \left( \frac{n}{2}\right)   \eu^{ - \frac{R^2+\|y\|^2}{2}}   }{ 2 \pi^{\frac{n}{2}}}    \frac{d}{d\|y\|}  \frac{   \mathsf{I}_{ \frac{n}{2}-1}(\|y\| R)}{ (\| y\| R)^{\frac{n}{2}-1}} \notag\\
%&=- \|y\| f_Y(y)  \notag\\
%&+  \frac{\Gamma \left( \frac{n}{2}\right)   \eu^{ - \frac{R^2+\|y\|^2}{2}}   }{ 2 \pi^{\frac{n}{2}}}  \left(    \frac{   R  \mathsf{I}_{ \frac{n}{2}-1}^\prime(\|y\| R)}{ (\| y\| R)^{\frac{n}{2}-1}} -  \frac{  \left(\frac{n}{2} -1\right)  R  \mathsf{I}_{ \frac{n}{2}-1}(\|y\| R)}{ (\| y\| R)^{\frac{n}{2}}} \right)\\
& = - \|y\| f_Y(y)  +  \frac{\Gamma \left( \frac{n}{2}\right)   \eu^{ - \frac{R^2+\|y\|^2}{2}}  R  }{ 2 \pi^{\frac{n}{2}}}  \notag\\
& \cdot \left(    \frac{     (\| y\| R) \mathsf{I}_{ \frac{n}{2}-1}^\prime(\|y\| R)- \left(\frac{n}{2} -1\right)    \mathsf{I}_{ \frac{n}{2}-1}(\|y\| R)}{ (\| y\| R)^{\frac{n}{2}}} \right)\\
&  \stackrel{a)}{=}- \|y\| f_Y(y)  +  \frac{\Gamma \left( \frac{n}{2}\right)   \eu^{ - \frac{R^2+\|y\|^2}{2}}   }{ 2 \pi^{\frac{n}{2}}}  \left(    \frac{   R^2 \|y\|  \mathsf{I}_{ \frac{n}{2}}(\|y\| R)  }{ (\| y\| R)^{\frac{n}{2}}} \right) \notag\\
&  \stackrel{b)}{=} - \|y\| f_Y(y)  +     \frac{   R  f_Y(y)  \mathsf{I}_{ \frac{n}{2}}(\|y\| R)  }{   \mathsf{I}_{ \frac{n}{2}-1}(\|y\| R) }  \notag\\
&  =- \|y\| f_Y(y)  +       R  f_Y(y)  \mathsf{h}_{ \frac{n}{2}}(\|y\| R)  , \label{eq:DerivativeNorm}
\end{align}
where the labeled equalities follow from: a) using the well-known recurrence relation $x\mathsf{I}_{v+1}(x) = x\mathsf{I}_v^\prime(x) - v\mathsf{I}_v(x)$ \cite{watson1995treatise}; and b) using the expression for $f_Y(y)$  in \eqref{eq:OuputDistribuiton}.   

The proof of \eqref{eq:ConditionalExpectation} is completed by combining  \eqref{eq:ScoreFunction}, \eqref{eq:ChainRule} and \eqref{eq:DerivativeNorm}.
\end{IEEEproof}

\begin{remark}
The proof of Proposition~\ref{prop:ConditionExpectation} relies on the following  identity between the conditional expectation and the output pdf  \cite{esposito1968relation}: 
\begin{align}
\E[X \mid Y=y]=y + \frac{ \nabla_y f_Y(y)}{f_Y(y)}, \label{eq:ScoreFunctionConditonalExpectation}
\end{align}
in which the quantity  $\frac{ \nabla_y f_Y(y)}{f_Y(y)}$ is commonly known as the score function. 
The application of the identity in \eqref{eq:ScoreFunctionConditonalExpectation} considerably simplifies the computation of $\E[X \mid Y]$ as we do not need to derive the conditional distribution $P_{X|Y}$ and only use properties of the output pdf   $f_Y(y)$. 
\end{remark}

Examples of shapes of the conditional expectation for $n=1$ and $n=2$ are shown on Fig.~\ref{fig:Estimators}.

\begin{figure}

\begin{subfigure}[a]{0.45\textwidth}
             \input{EstimatorN1.tex}
\caption{Case of $n=1$. The conditional expectation $\E \left[  X_R | Y =y \right]=R \tanh(Ry)$  where $R=2$. }
\label{eq:condExpectn=1}
        \end{subfigure}%

\begin{subfigure}[c]{0.45\textwidth}
              \input{Estimator.tex}
%\caption{Case of $n=2$. The conditional expectation $\E\left[  X_1 | Y =y \right]$  where $X_R=[X_1,X_2]$ with  $R=2.5$ and where $y=[y_1,y_2]$. }
        \end{subfigure}%
        
        \caption{Examples  of the conditional expectation in \eqref{eq:ConditionalExpectation} for $n=1$ and $n=2$.}
\label{fig:Estimators}
\end{figure}

The mutual information and the MMSE of $X_R$ are given next. 

\begin{prop} \label{prop:MianMMSE} \emph{(MMSE and Mutual Information)} For every $R>0$
\begin{align}
I(X_R;Y)&=R^2 \log(\eu)+ \log \left( \frac{2^{1-\frac{n}{2}}}{\Gamma \left( \frac{n}{2} \right)} \right)  \notag\\
& - \E\left[  \log \left(  \frac{  \mathsf{I}_{ \frac{n}{2}-1}(\|Z+x\| R)}{ (\| Z+x\| R)^{\frac{n}{2}-1}} \right)  \right], \label{eq:Mutualinfor}
\end{align}
and
\begin{align}
\mmse(X_R \mid Y)= R^2 -  R^2 \E \left[    \mathsf{h}_{\frac{n}{2}}^2\left(  R   \| x+ Z \| \right) \right], \label{eq:MMSEformula}
\end{align} 
for any $\|x\|=R$. 
\end{prop}
\begin{IEEEproof}
First observe that due to the symmetry  of $i(x,P_{X_R})$ and $X_R$ we have that
\begin{align*}
I(X_R;Y)= \E[  i(X,P_{X_R})] = i( x,P_{X_R}),
\end{align*}
where $\|x\|=R$. Therefore, 
\begin{align*}
&I(X_R;Y)+ h(Z)\\
&= \int_{\mathbb{R}^n}  \frac{1}{ (2 \pi)^{\frac{n}{2}}}  \eu^{-\frac{\| y-x\|^2}{2}} \log \frac{1}{f_Y(y)} {\rm d}y \\
&= \log \left( \frac{2 \pi^{\frac{n}{2}}}{\Gamma \left( \frac{n}{2} \right)} \right) + \frac{R^2+\E[\| Z+x\|^2] }{2}\log(\eu) \notag\\  
&+ \E\left[  \log \left(  \frac{ (\| Z+x\| R)^{\frac{n}{2}-1}}{  \mathsf{I}_{ \frac{n}{2}-1}(\|Z+x\| R)} \right)  \right]\\
&=\log \left( \frac{2  (\pi \eu)^{\frac{n}{2}}}{\Gamma \left( \frac{n}{2} \right)} \right)  + R^2 \log(\eu) \notag\\
&  + \E\left[  \log \left(  \frac{ (\| Z+x\| R)^{\frac{n}{2}-1}}{  \mathsf{I}_{ \frac{n}{2}-1}(\|Z+x\| R)} \right)  \right]. 
\end{align*} 
This concludes the proof of \eqref{eq:Mutualinfor}. To show \eqref{eq:MMSEformula} observe that the MMSE can be written as
\begin{align*}
\mmse(X_R|Y)&= \E[\|X_R\|^2] -  \E \left[ \| \E[X_R \mid Y] \|^2 \right]\\
&=R^2-   \E \left[ \| \E[X_R \mid Y] \|^2  \mid \|X_R\|=R \right]\\
&=R^2-     R^2 \E \left[    \mathsf{h}_{\frac{n}{2}}^2\left(  R   \| x+ Z \| \right) \right],
\end{align*} 
where $\|x\|=R$. 
This concludes the proof. 
\end{IEEEproof}

\begin{figure}
\input{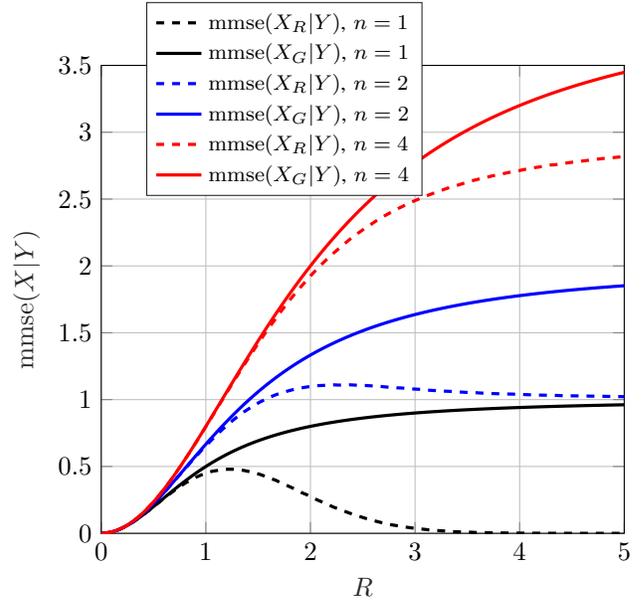}
\caption{Comparison of the MMSE of $X_R$  (dashed line)  and  the MMSE of  $X_G \sim \mathcal{N}(0, R^2 I_n)$ (solid line).   }
\label{fig:MMSEcurves}
\end{figure}

Fig.~\ref{fig:MMSEcurves} shows the MMSE of  $X_R$  in \eqref{eq:MMSEformula} vs. the MMSE of $X_G \sim \mathcal{N}(0, R^2 I_n)$ which is given by
\begin{align}
\mmse(X_G|Y)=  n \frac{ \frac{1}{n} R^2}{1+ \frac{1}{n} R^2}. 
\end{align}

%\section{Asymptotic}
%
%
%
%First question is what is 
%\begin{align}
%\lim_{R \to \infty}  I(X;Y) 
%\end{align}
%
%
%To answer this we must find
%\begin{align}
% \lim_{R \to \infty} \E\left[  \log \left(  \frac{  I_{ \frac{n}{2}-1}(\|Z+x\| R)}{  \eu^{R^2 }(\| Z+x\| R)^{\frac{n}{2}-1}} \right)  \right]
%\end{align}
%
%  
%  
%{\color{red} NEED good bounds} 
%  

\section{ A  New Condition for Optimality in the Small Amplitude Regime} 
\label{sec:NewCondition}

In this section an equivalent  optimality condition  to that in Theorem~\ref{thm:OptimalityCondition} is derived.  The new condition has the advantage of being  easier to verify than the condition in Theorem~\ref{thm:OptimalityCondition}. 

The following two lemmas would be useful in our analysis.  
\begin{lem}\label{lem:OrthogonalEquvariance} The function $ x \mapsto i(x, P_{X_R})$ is  a function only of $\|x\|$. 
\end{lem}
\begin{IEEEproof}
The proof follows from the symmetry of the Gaussian distribution and the symmetry of $P_{X_R}$.
\end{IEEEproof}

The next lemma was shown in \cite[Theorem 3]{karlin1957polya}. 
\begin{lem} \label{lem:numberOfOscillations} Let the pdf $f(x,\omega)$ be a positive-definite kernel that can be differentiated $n$ times with respect to $x$ for all $\omega$, and let $\eta(\omega)$ be a function that changes  sign $n$ times. If
\begin{align}
M(x)\coloneqq \int    \eta(\omega)     f(x,\omega) {\rm d} \omega,
\end{align} 
can be differentiated $n$ times, then $M(x)$ changes sign at most $n$ times. 
\end{lem}

\begin{theorem} \emph{(A New Optimality Condition)}\label{thm:NewCondition} $P_{X_R}$   is optimal if and only if  for all $\|x\|=R$
\begin{align}
i(0, P_{X_R}) \le i(x, P_{X_R}), \label{eq:newCOndiitonForOptimality}.
\end{align}
\end{theorem}
\begin{IEEEproof}
Since by Lemma~\ref{lem:OrthogonalEquvariance} $i(x, P_{X_R})$ is  a function  only of $\|x\|$ let
\begin{align}
g(\|x \|)\coloneqq i(x, P_{X_R}).
\end{align}
The goal now is to show that  the maximum of  $g(\|x \|)$  for $x \in \mathcal{B}_0(R)$ occurs either at $\|x\|=0$ or $\|x\|=R$. This would simplify the two conditions in \eqref{eq:equalityCondition} and \eqref{eq:inequalitityCondition} to only one  condition
\begin{align}
g(0) \le g(R). 
\end{align}
In order to show this claim, we prove that the derivative of $g(\|x \|)$ makes only one sign change, and that sign change is from negative to positive. Hence, $g(\|x \|)$ has only one local minimum and must be maximized only  at the boundaries $\|x\|=0$ and $\|x\|=R$.

Because $g(\|x \|)$  depends on $x$ only through $\|x\|$, there is no loss of generality in taking $x=[x_1,0,...,0]$. Consider the derivative of $g(x_1)$ with respect to $x_1$
\begin{align*}
&g^\prime(x_1) \notag\\%& = \frac{d}{dx_1}  \int_{\mathbb{R}^n}  \frac{1}{ (2 \pi)^{\frac{n}{2}}}  \eu^{-\frac{\| y -x \|^2}{2}} \log \frac{1}{f_Y(y)} {\rm d}y\\
&=  \frac{d}{dx_1}  \int_{\mathbb{R}^n}  \frac{1}{ (2 \pi)^{\frac{n}{2}}}  \eu^{- \frac{\sum_{i=2}^n y_i^2 + (y_1-x_1)^2}{2}} \log \frac{1}{f_Y(y)} {\rm d}y\\
&=   \int_{\mathbb{R}^n}  \frac{1}{ (2 \pi)^{\frac{n}{2}}}  \eu^{- \frac{\sum_{i=2}^n y_i^2 + (y_1-x_1)^2}{2}}  (y_1-x_1) \log \frac{1}{f_Y(y)} {\rm d}y\\
&= -   \int_{\mathbb{R}^n}  \frac{1}{ (2 \pi)^{\frac{n}{2}}}  \eu^{- \frac{\sum_{i=2}^n y_i^2 + (y_1-x_1)^2}{2}}  (y_1-x_1) \log f_Y(y) {\rm d}y.
\end{align*}

Integrating by parts with respect to $y_1$ we have that
\begin{align*}
g^\prime(x_1) &=    -\int_{\mathbb{R}^n}  \frac{1}{ (2 \pi)^{\frac{n}{2}}}  \eu^{- \frac{\sum_{i=2}^n y_i^n + (y_1-x_1)^2}{2}}   \rho(y) {\rm d}y,
\end{align*}
where 
\begin{align*}
  \rho(y)&\coloneqq \frac{d}{dy_1}  \log f _Y(y)=  \frac{  \frac{d}{dy_1} f_Y(y)}{f_Y(y)}.
\end{align*}
Next using the  chain rule of differentiation we have that 
\begin{align*}
  \frac{d}{dy_1} f_Y(y) &=    \frac{d}{d\|y\|} f_Y(y)  \frac{y_1}{\|y\|}\\
  & = \left(- \|y\| f_Y(y)  +     \frac{   R  f_Y(y)  \mathsf{I}_{ \frac{n}{2}}(\|y\| R)  }{   \mathsf{I}_{ \frac{n}{2}-1}(\|y\| R) } \right) \frac{y_1}{\|y\|},
\end{align*}
where in the last step we have used \eqref{eq:DerivativeNorm}.  Therefore,
\begin{align*}
  \rho(y)&= \left(- \|y\|  +     R    \mathsf{h}_{\frac{n}{2}}(\|y\|R)  \right) \frac{y_1}{\|y\|}\\
  &\coloneqq M(\|y\|)  \frac{y_1}{\|y\|}.
\end{align*}
Next by transforming to spherical coordinates we have that 
\begin{align}
g^\prime(x_1)& =   -   2 x_1 \int_{0}^\infty    M(r)   \eu^{- \frac{r^2+ x_1^2}{2}} \frac{1}{2}\left( \frac{r}{x_1 } \right)^{\frac{n}{2}}  \mathsf{I}_{\frac{n}{2}}(x_1 r )   {\rm d}r  \label{eq:aklfnlkandf;al}\\
& =     - 2 x_1  \E \left[ M( \sqrt{V^2})  \right], \label{eq:jalkfnalnflkadflk}
\end{align}
where $V^2$ is the non-central chi-square distribution with $n+2$ degrees of freedom and non-centrality $x_1^2$; see Appendix~\ref{sec:SphericalCoordinates} for the derivation \eqref{eq:aklfnlkandf;al} and \eqref{eq:jalkfnalnflkadflk}.

Another fact which is not difficult to check is that  for  large enough $x_1$   the function    $g^\prime(x_1)$ is positive.   This is shown next
\begin{align}
 - 2 x_1  \E \left [ M \left( \sqrt{V^2} \right) \right] &=  - 2 x_1  \E \left[ -  \sqrt{V^2}   +     R    \mathsf{h}_{\frac{n}{2}} \left( \sqrt{V^2} R \right) \right ] \notag\\
 &= 2 x_1 \left(   \E \left[  \sqrt{V^2}  \right]-  \E \left[    R    \mathsf{h}_{\frac{n}{2}}( \sqrt{V^2} R)  \right] \right)\notag\\
    &\stackrel{a)}{\ge}  2 x_1 \left(    \sqrt{\E[ V^2 ]} -   R  \right )\notag\\
       &\stackrel{b)}{\ge}  2 x_1 \left(     \sqrt{n+2 +x_1^2} -   R  \right), \label{eq:BoundOnEofM}
\end{align}
where the labeled (in)-equalities follow from: a) using   $\mathsf{h}_{\frac{n}{2}}( \sqrt{V^2} R) \le 1$ and  $\E \left[  \sqrt{V^2} \right]  \ge  \sqrt{\E[ V^2 ]}$; and b) using  the  expression of the mean of  the non-central chi-square distribution with $n+2$ degrees of freedom and non-centrality $x_1^2$.  Therefore,   in view of the bound in \eqref{eq:BoundOnEofM},  the expression in \eqref{eq:jalkfnalnflkadflk} is positive for  $x_1$  large enough.

Next observe that in \eqref{eq:aklfnlkandf;al}   the function
\begin{align*}
M(r)= - r +        R    \mathsf{h}_{ \frac{n}{2}}(r R) ,
\end{align*}
changes sign at most once for $r>0$, which follows from the fact that $ \mathsf{h}_{ \frac{n}{2}}(x) $ is increasing and concave (see \cite{watson1983statistics}) and $\mathsf{h}_{ \frac{n}{2}}(0)=0$.  Hence, using Lemma~\ref{lem:numberOfOscillations} we have that for $x_1>0$  the function  $g^\prime(x_1)$ changes sign at most once, and since $g^\prime(x_1)>0$  for large enough $x_1$,  we conclude that  the sign change can only be  from negative to positive.  Therefore,  for $x_1 >0$ the function  $g(x_1)$ has only one local minimum, no local maxima, and $g(\|x\|)$ is  maximized only at the boundaries.  
This concludes the proof. 
\end{IEEEproof}

\begin{remark}
Condition \eqref{eq:newCOndiitonForOptimality} significantly simplifies the necessary and sufficient conditions for  optimality in \eqref{eq:SuffAndNecc}.  For instance, we do not have to verify the conditions in \eqref{eq:inequalitityCondition} for all $x \in \mathcal{B}_0(R)$ and instead need only to check the points satisfying $\|x\|=0$ and $\|x\|=R$.  

Moreover, the condition in \eqref{eq:newCOndiitonForOptimality} implies that as we increase  $R$ the new points of support cannot appear for $0<\| x\| <R$  and shows that a new probability  mass, as we transition beyond $\bar{R}_n$, can only appear at $\|x\|=0$. 
\end{remark}

\begin{figure}

\begin{subfigure}[a]{0.45\textwidth}
\center
             \input{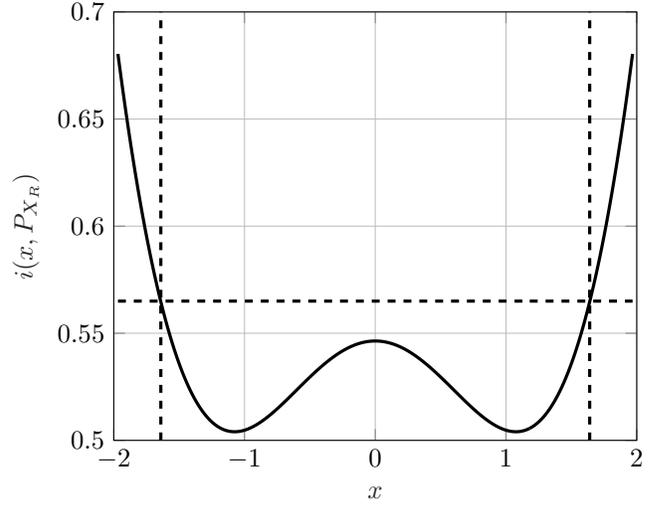}
\caption{Plot of $i(x,P_{X_R})$ vs. $x$ for $R=1.64$.}
        \end{subfigure}%
        ~
        
\begin{subfigure}[b]{0.45\textwidth}
\center
              \input{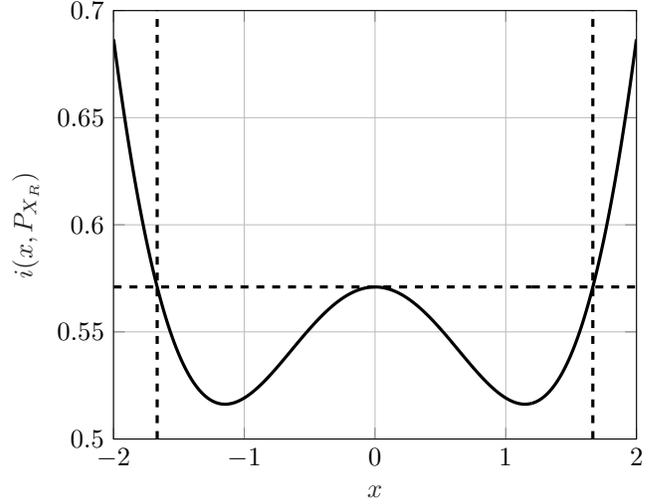}
              \caption{Plot of $i(x,P_{X_R})$ vs. $x$ for $R=\bar{R}_1 =1.665925641$. }
              \vspace{0.3cm}
        \end{subfigure}%
        ~
        
        \begin{subfigure}[b]{0.45\textwidth}
\center
              \input{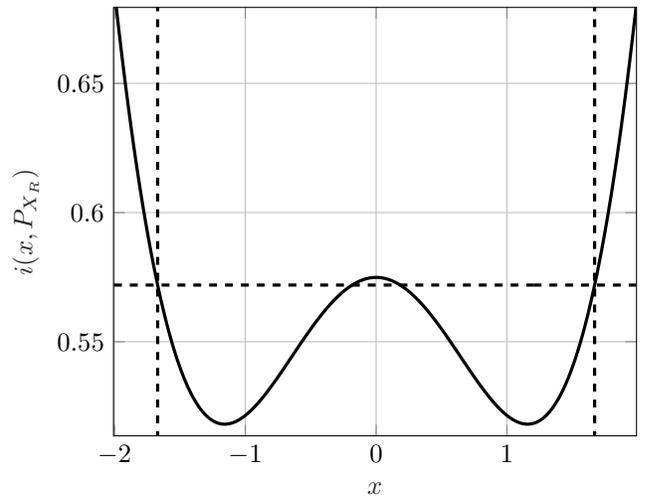}
              \caption{Plot of $i(x,P_{X_R})$ vs. $x$ for $R=1.67$.}
        \end{subfigure}%

\caption{Plot of $i(x,P_{X_R})$ vs. $x$ for $n=1$.  Solid lines are $i(x,P_{X_R})$ and vertical dashed lines are $x=R$ and $x=-R$.  }
\label{fig:InfoDensity}
\end{figure}

Fig.~\ref{fig:InfoDensity} shows $i(x,X_R)$ vs.  $x$ for $n=1$.  Note that, as expected,  when $R=\bar{R}_1$ we have that  $i(0,P_{X_R})=i(R,P_{X_R})$. Moreover, for $R>\bar{R}_1$ as $X_R$ is no longer optimal, we have that $i(0,P_{X_R})>i(R,P_{X_R})$

Next, we rewrite $i(0,P_{X_R})$ and $i(x,P_{X_R})$ in terms of  estimation theoretic measures which facilitates the computation of $\bar{R}_n$. 
\begin{lem}\label{lem:MMSEandMI} For every $R>0$ and $\|x\|=R$
\begin{align}
i(x,P_{X_R})&=  \frac{1}{2} \int_{0}^1 \hspace{-0.06cm}   \E \left[ \left\| X_R -\E[X_R \mid Y_\gamma] \|^2 \mid \|X_R \right\|=R\right]   {\rm d} \gamma,  \label{eq:MutualInfoA}  \\
i(0,P_{X_R})&= \frac{1}{2}  \int_{0}^1  \hspace{-0.06cm}  \E \left[  \left \| X_R -\E[X_R \mid Y_\gamma]  \right\|^2  \mid \|X_R\|=0 \right] {\rm d} \gamma,  \label{eq:MutalInfo0}
\end{align}
where  $Y_\gamma=\sqrt{\gamma}X_R+Z$. 
\end{lem}
\begin{IEEEproof}
The proof of   \eqref{eq:MutualInfoA}  follows by a symmetry argument used in  Proposition~\ref{prop:MianMMSE} and the I-MMSE relationship \cite{I-MMSE}
\begin{align}
I(X;Y)= \frac{1}{2}\int_{0}^1 \E \left[\| X -\E[X \mid Y_\gamma] \|^2 \right] {\rm d} \gamma. 
\end{align}

To show \eqref{eq:MutalInfo0} we use the point-wise I-MMSE formula  \cite{pointWiseI-MMSE} 
\begin{align}
&\log  \frac{f_{Y|X}(Y|X)}{ f_{Y}(Y)} - \frac{1}{2}\int_0^1 \| X-\E[X \mid Y_\gamma]\|^2 {\rm d} \gamma \notag\\
& =   \int_0^1 (X-   \E[X  \mid Y_\gamma]) \cdot {\rm d} W_\gamma,  \, \text{a.s.}\label{eq:PointWiseI-MMSE}
\end{align}
where the integral on the right hand side of \eqref{eq:PointWiseI-MMSE} is the It\^o integral  with respect to $W_\gamma$.   The proof of the representation of $i(0,P_{X_R}) $ now goes as follows: 
\begin{align*}
2 i(0,P_{X_R}) 
&= \E \left[  \log  \frac{f_{Y|X_R}(Y|X_R)}{ f_{Y}(Y)}  \mid \|X_R\|=0 \right]\\
& \stackrel{a)}{=}  \E \left[   \int_0^1 \|X_R-\E[X_R \mid Y_\gamma]\|^2 {\rm d} \gamma  \mid \|X_R\|=0  \right] \notag\\
&-  \E \left[   \int_0^1 (X_R-   \E[X_R \mid Y_\gamma]) \cdot {\rm d} W_\gamma   \mid X_R=0 \right]\\
& \stackrel{b)}{=}   \E \left[   \int_0^1 \|X_R-\E[X_R \mid Y_\gamma]\|^2 {\rm d} \gamma  \mid X_R=0 \right]\\
&=   \int_0^1  \E \left[   \|X_R-\E[X_R \mid Y_\gamma]\|^2    \mid  X_R=0 \right] {\rm d} \gamma ,
\end{align*}
where the labeled equalities follow from: a) using the point-wise formula in \eqref{eq:PointWiseI-MMSE}; and b) using the symmetry of $X_R$ to conclude that $\E[ \E[X_R|Y_\gamma] | X_R=0] =0$. This concludes the proof. 
\end{IEEEproof}

\subsection{Proof of Theorem~\ref{thm:MainResult}} 
\label{sec:ProofOfTheMainResult}
Combining Lemma~\ref{lem:MMSEandMI} and the optimality condition in Theorem~\ref{thm:NewCondition} we arrive at
\begin{align}
0&\ge i(0,P_{X_R})-i(x,P_{X_R})   \notag\\
&=   \int_{0}^1   \E \left[\| X_R -\E[X_R \mid Y_\gamma] \|^2  \mid \| X_R\|=0\right]  \notag\\ 
&- \E \left[\| X_R -\E[X_R \mid Y_\gamma] \|^2  \mid  \|X_R\|=R\right]   {\rm d} \gamma \notag\\
&=   \int_{0}^1   R^2 \E \left[    \mathsf{h}_{\frac{n}{2}}^2\left(  \sqrt{\gamma} R   \|  Z \| \right) \right] - R^2  \notag\\
&- R^2 \E \left[    \mathsf{h}_{\frac{n}{2}}^2\left(  \sqrt{\gamma} R   \|  \sqrt{\gamma} x+ Z \| \right) \right]  {\rm d} \gamma,  \label{eq:ConditionForOptimalityMMSE}
\end{align}
where in the last step we have used the expression for the conditional expectation in \eqref{eq:ConditionalExpectation} and the expression for the MMSE in \eqref{eq:MMSEformula}.
Now the condition in \eqref{eq:ConditionForOptimalityMMSE} is equivalent to 
\begin{align}
   \int_{0}^1   \hspace{-0.05cm} \E \left[    \mathsf{h}_{\frac{n}{2}}^2\left(  \sqrt{\gamma} R   \|  Z \| \right) \right] + \E \left[    \mathsf{h}_{\frac{n}{2}}^2\left(  \sqrt{\gamma} R   \|  \sqrt{\gamma}  x+ Z \| \right) \right]  {\rm d} \gamma \le 1 \label{eq:ConditionForTheOptimalityOfSphere},
\end{align} 
where $\|x\|=R$.
The value of  $\bar{R}_n$ would now be  a solution of \eqref{eq:ConditionForTheOptimalityOfSphere} which concludes the proof of \eqref{eq:ExactR}. 

To show the second part of Theorem~\ref{thm:MainResult}  let  $R=c \sqrt{n}$. We will also need   the following bounds   on $\mathsf{h}_v(x)$ \cite{segura2011bounds,baricz2015bounds}:
\begin{align}
\mathsf{h}_v(x) &\ge \frac{x}{ v  +\sqrt{ v^2  +x^2}},  \text{ for }  v>0,\\
\mathsf{h}_v(x) &\le  \frac{x}{  \frac{2v-1}{2}  +\sqrt{  \frac{ \left(2v-1\right)^2 }{4}  +x^2}},  \text{ for }  v>\frac{1}{2}.
\end{align} 
Moreover, if we let $\bar{ x}=[c \sqrt{n},0,0,...]$ and  define
\begin{align}
 V_n &\coloneqq  \frac{1}{\sqrt{n}} \|  cZ +  c  \sqrt{\gamma} \bar{x}\|,\\ 
  W_n &\coloneqq  \frac{1}{\sqrt{n}} \|  cZ \|,
\end{align}
then   the two terms on the  left hand side of \eqref{eq:ConditionForTheOptimalityOfSphere} can be lower and upper bounded as follows:
\begin{align}
%   \E \left[    \left( \frac{  c \sqrt{n}    \|  \Z \| }{ \frac{n}{2}+\sqrt{ \frac{n^2}{4} + c^2 n    \|  \Z \|^2}}  \right)^2 \right] &\le  \E \left[    \mathsf{h}_{\frac{n}{2}}^2\left( R   \| \Z \| \right) \right]  \le \E \left[    \left( \frac{  c \sqrt{n}    \|  \Z \| }{ \frac{n-1}{2}  +\sqrt{ \frac{(n-1)^2}{4}  + c^2 n    \|  \Z \|^2}}  \right)^2 \right] \label{eq:ajfklafna;lkndl}\\
%   
&\E \left[    \left( \frac{  \sqrt{\gamma} W_n }{ \frac{1}{2 }  +\sqrt{ \frac{1}{4}  + \gamma W_n^2}}  \right)^2 \right] \le    \E \left[    \mathsf{h}_{\frac{n}{2}}^2\left( c \sqrt{n} \sqrt{ \gamma}   \| Z \| \right) \right]  \notag\\
&\le      \E \left[    \left( \frac{   \sqrt{\gamma} W_n }{ \frac{n-1}{2 n}  +\sqrt{ \frac{(n-1)^2}{4 n^2}  + \gamma W_n^2}}  \right)^2 \right] , \label{eq:ajfklafna;lkndl}
\end{align}
and
\begin{align}
& \E \left[    \left( \frac{  \sqrt{\gamma} V_n }{ \frac{1}{2 }  +\sqrt{ \frac{1}{4}  + \gamma V_n^2}}  \right)^2 \right] \le    \E \left[    \mathsf{h}_{\frac{n}{2}}^2\left( c \sqrt{n}  \sqrt{ \gamma}    \|  \sqrt{\gamma} \bar{x}+ Z \| \right) \right]  \notag\\
&\le      \E \left[    \left( \frac{  \sqrt{\gamma} V_n }{ \frac{n-1}{2 n}  +\sqrt{ \frac{(n-1)^2}{4 n^2}  +\gamma V_n^2}}  \right)^2 \right] , \label{eq:jaklfnalnfd;landflkan}
%      \E \left[    \left( \frac{  c \sqrt{n}     \| \bar{{\bf x}}+ \Z \| }{ \frac{n}{2} +\sqrt{\frac{n^2}{4} + c^2 n    \| \bar{{\bf x}}+ \Z \|^2}}  \right)^2 \right] &\le  \E \left[    \mathsf{h}_{\frac{n}{2}}^2\left( R    \| \bar{{\bf x}}+ \Z \|\right) \right]  \le \E \left[    \left( \frac{  c \sqrt{n}    \| \bar{{\bf x}}+ \Z \| }{ \frac{n-1}{2}  +\sqrt{ \frac{(n-1)^2}{4}  + c^2 n    \| \bar{{\bf x}}+ \Z \|^2}}  \right)^2 \right] \label{eq:jaklfnalnfd;landflkan}
\end{align} 
where the lower bounds hold for $n \ge 1$ and the upper bounds hold for $n>1$.

   In view of the fact that $ u \mapsto  \frac{u}{(a+\sqrt{a^2+u})^2}$ is a concave function for $a>0$, using Jensen's inequality, we can further upper bound the expressions in \eqref{eq:ajfklafna;lkndl} and \eqref{eq:jaklfnalnfd;landflkan} as follows:
\begin{align}
 &\E \left[    \left( \frac{  \sqrt{\gamma} W_n }{ \frac{n-1}{2 n}  +\sqrt{ \frac{(n-1)^2}{4 n^2}  + \gamma W_n^2}}  \right)^2 \right] \notag\\
  &\le     \frac{  \gamma  \E[ W_n^2] }{ \left(\frac{n-1}{2 n}  +\sqrt{ \frac{(n-1)^2}{4 n^2}  + \gamma \E[ W_n^2]} \right)^2}  \notag\\
 &= \frac{   \gamma c^2 }{ \left( \frac{n-1}{2 n}  +\sqrt{ \frac{(n-1)^2}{4 n^2}  + \gamma c^2} \right)^2},  \label{eq:akljfalnfajkfdajkhf}
 \end{align}
 and
 \begin{align}
 & \E \left[    \left( \frac{  \sqrt{\gamma}V_n }{ \frac{n-1}{2 n}  +\sqrt{ \frac{(n-1)^2}{4 n^2}  + \gamma^2 V_n^2}}  \right)^2 \right]  \notag\\
 &\le     \frac{  \gamma  \E[ V_n^2] }{ \left( \frac{n-1}{2 n}  +\sqrt{ \frac{(n-1)^2}{4 n^2}  +  \gamma  \E[ V_n^2]} \right)^2}  \notag\\
  &= \frac{    \gamma c^2 (1+\gamma c^2) }{ \left( \frac{n-1}{2 n}  +\sqrt{ \frac{(n-1)^2}{4 n^2}  +   \gamma c^2 (1+ \gamma c^2)} \right)^2}, \label{eq:jalkdjfalndfandfhadha}
 \end{align}
 where we have also used that  $   \E[ W_n^2]=c^2$ and $\E[ V_n^2]= c^2 (1+\gamma c^2)$.  Applying the bounds in  \eqref{eq:akljfalnfajkfdajkhf} and \eqref{eq:jalkdjfalndfandfhadha}   to a necessary and sufficient condition in \eqref{eq:ConditionForTheOptimalityOfSphere} we arrive at the following sufficient condition for optimality: 
 \begin{align}
&\int_0^1 \frac{ \gamma  c^2 }{ \left( \frac{n-1}{2 n}  +\sqrt{ \frac{(n-1)^2}{4 n^2}  + \gamma c^2} \right)^2}  \notag\\
&+\frac{  \gamma c^2(1+\gamma c^2) }{ \left( \frac{n-1}{2 n}  +\sqrt{ \frac{(n-1)^2}{4 n^2}  + \gamma c^2(1+ \gamma c^2)} \right)^2} {\rm d} \gamma \le 1. \label{eq:djflakndfliiiii}
 \end{align} 
 
 Next, we verify that it is sufficient to take $R \le \sqrt{n}$ which is equivalent to verifying that the inequality in \eqref{eq:djflakndfliiiii}  holds for $c=1$. Choosing $c=1$ in \eqref{eq:djflakndfliiiii}  and letting $a=\frac{n-1}{2 n}$ we arrive at the following inequality:
  \begin{align}
%&6\,a^2-2\,a\,\sqrt{a^2+2}-6\,a^2\,\ln\left(2\,a\right)-4\,a\,\sqrt{a^2+1} \notag\\
%&+4\,a^2\ln\left(a+\sqrt{a^2+1}\right)+2a^2\ln\left(a+\sqrt{a^2+2}\right)+2 \le 1. 
&2 a \log(2a+1) - 2a \log \left( 2 \sqrt{a^2+2}+3 \right) \notag\\
&-4a^2\tanh^{-1}\left(8 a^2-1\right)+ 4 a^2 \tanh^{-1} \left(  \frac{4a \sqrt{a^2+2}-1}{3}\right) \notag\\
&\le 1.  \label{eq:flkadlkfnanlajfa}
\end{align}
The solution to the above inequality can be found numerically and  is given by $   0.2358 \le a=\frac{n-1}{2 n}$    or   $n \ge  1.892$. 
Therefore, for $n \ge 2$ it is sufficient to take $c=1$ or $R \le \sqrt{n}$.

 Next, we find the exact limiting behavior of $c$. Observe that the lower and upper bounds in \eqref{eq:ajfklafna;lkndl} and \eqref{eq:jaklfnalnfd;landflkan} are  equal as  $n \to \infty$ and, therefore, we  focus  only on the upper bounds in \eqref{eq:ajfklafna;lkndl} and \eqref{eq:jaklfnalnfd;landflkan}. By the strong law of large numbers almost surely we have the following limits:
\begin{align}
\lim_{n \to \infty}  V_n^2&=   \lim_{n \to \infty}   \frac{1}{n} \|  cZ +  c \sqrt{\gamma} \bar{x}\|^2 \notag\\
&=  \lim_{n \to \infty}  \frac{1}{n}  (cZ_1+c^2 \sqrt{\gamma} \sqrt{n})^2 +  \lim_{n \to \infty}  \frac{1}{n}  \sum_{i=2}^n  (cZ_i)^2\notag\\
&=  c^2(1+\gamma c^2),
\end{align}
and similarly
\begin{align}
  \lim_{n \to \infty}W_n^2&=   \lim_{n \to \infty}  \frac{1}{n}  \sum_{i=1}^n  (cZ_i)^2=c^2.
\end{align}

Now to show that the limit and the expectation can be interchanged observe that 
\begin{align}
\left| \frac{  \sqrt{\gamma}V_n }{ \frac{n-1}{2 n}  +\sqrt{ \frac{(n-1)^2}{4 n^2}  +\gamma V_n^2}}  \right|\le 1,\\
\ \left|\frac{  \sqrt{\gamma}W_n }{ \frac{n-1}{2 n}  +\sqrt{ \frac{(n-1)^2}{4 n^2}  +\gamma W_n^2}} \right| \le 1,
\end{align}
and  by the Dominated Convergence Theorem the limits are give by 
\begin{align}
   & \lim_{ n \to \infty} \E \left[    \left( \frac{  c \sqrt{n}  \sqrt{\gamma}   \|  Z+ \sqrt{\gamma} \bar{ x} \| }{ \frac{n-1}{2}  +\sqrt{ \frac{(n-1)^2}{4}  + c^2 n  \gamma   \|  Z+ \sqrt{\gamma}  \bar{ x} \|^2}}  \right)^2 \right] \notag\\
     &=    \lim_{ n \to \infty}  \E \left[    \left( \frac{  \sqrt{\gamma} V_n }{ \frac{n-1}{2 n}  +\sqrt{ \frac{(n-1)^2}{4 n^2}  + \gamma V_n^2}}  \right)^2 \right] \notag\\
&=       \left( \frac{  \sqrt{\gamma} c \sqrt{1+ \gamma c^2} }{ \frac{1}{2} +\sqrt{ \frac{1}{4} +  \gamma c^2(1+ \gamma c^2)}}  \right)^2, \label{eq:jaklfalfnalkddd}
\end{align} 
and 
\begin{align}
 &  \lim_{ n \to \infty} \E \left[    \left( \frac{  c \sqrt{n}  \sqrt{\gamma}  \| Z \| }{ \frac{n-1}{2}  +\sqrt{ \frac{(n-1)^2}{4}  + c^2 n   \sqrt{\gamma}  \| Z \|^2}}  \right)^2 \right]  \notag\\
&=    \lim_{ n \to \infty} \E \left[    \left( \frac{ \sqrt{\gamma}  W_n }{ \frac{n-1}{2 n}  +\sqrt{ \frac{(n-1)^2}{4 n^2}  + \gamma W_n^2}}  \right)^2 \right] \notag\\
&=       \left( \frac{  \sqrt{\gamma}  c }{ \frac{1}{2} +\sqrt{ \frac{1}{4} + \gamma c^2}}  \right)^2. \label{eq:nnfahfahgppp}
\end{align}

Therefore, the condition for optimality is given by 
  \begin{align}
\int_0^1  \hspace{-0.16cm} \frac{ \gamma  c^2 }{ \left( \frac{1}{2 }  +\sqrt{ \frac{1}{4 }  \hspace{-0.05cm} + \hspace{-0.05cm} \gamma c^2} \right)^2}    \hspace{-0.05cm}  +\frac{  \gamma c^2(1+ \gamma c^2) }{ \left( \frac{1}{2 }  \hspace{-0.05cm}  +\sqrt{ \frac{1}{4 }  + \gamma c^2(1+ \gamma c^2)} \right)^2} {\rm d} \gamma =1. \label{eq:djflakndfliiiiiajfakljflka}
 \end{align}

\begin{figure*}[h!]
\begin{align}
%&\frac{\log\left(\sqrt{c^2\,\left(c^2+1\right)+\frac{1}{4}}+\frac{1}{2}\right)-2\,\sqrt{c^2\,\left(c^2+1\right)+\frac{1}{4}}+1}{c^2\,\left(c^2+1\right)}-\frac{\log\left(2\right)-\log\left(2\,\sqrt{c^2+\frac{1}{4}}+1\right)+2\,\sqrt{c^2+\frac{1}{4}}-1}{c^2}+2=1 \label{eq:ExpressionForTheIntegration}\\
\frac{\log\left(\sqrt{4c^2+1}+1\right)-\log\left(c^2+1\right)-\log\left(2\right)-\sqrt{4c^2+1}+2\,c^2+1}{c^2}=1 \label{eq:ExpressionForTheIntegration}
\end{align}
\end{figure*}

The integral in \eqref{eq:djflakndfliiiiiajfakljflka} does have a closed form expression given in \eqref{eq:ExpressionForTheIntegration}, however, the resulting equation must be solved numerically. Using numerical methods it is not difficult to verify that the solution to the equation in \eqref{eq:djflakndfliiiiiajfakljflka} is given by  $c \approx 1.860935682$.  This concludes the proof.

\subsection{Proof of Theorem~\ref{thm:AnotherAltertnativeCondtion}}
\label{sec:proof:thm:AnotherAltertnativeCondtion}
First we compute the difference $i(x, P_X)- i(0, P_X) $ where we take $x=[x_1,0,...,0]$ 
\begin{align}
&i(x, P_X)- i(0, P_X) \notag\\
&= \int_{\mathbb{R}^n}  \left(  \frac{1}{ (2 \pi)^{\frac{n}{2}}}  \eu^{-\frac{\| y-x\|^2}{2}} -   \frac{1}{ (2 \pi)^{\frac{n}{2}}}  \eu^{-\frac{\| y\|^2}{2}}  \right) \log \frac{1}{f_Y(y)} {\rm d}y. \label{Eq:diffofDenistities}
\end{align} 
Next considering only  the integral with respect to $y_1$, we have
\begin{align}
&-\int_{\mathbb{R}}  \left(  \frac{1}{  \sqrt{2 \pi} } \eu^{-\frac{ (y_1-x_1)^2}{2}} -   \frac{1}{ \sqrt{2 \pi} }  \eu^{-\frac{y_1^2}{2}}  \right) \log f_Y(y) {\rm d}y_1\\
&\stackrel{a)}{=}\int_{\mathbb{R}}  \left( Q(y_1)-    Q(y_1-x_1)  \right)  \frac{\frac{d}{dy_1} f_Y(y) }{ f_Y(y) }  {\rm d}y_1 \notag\\
& \stackrel{b)}{=} \int_{\mathbb{R}}  \left( Q(y_1)-    Q(y_1-x_1)  \right)   \left( \E[X_1  \mid Y=y]- y_1 \right) {\rm d}y_1 \notag\\
&  \stackrel{c)}{=}  \int_{\mathbb{R}}  \left( Q(y_1)-    Q(y_1-x_1)  \right)   \E[X_1  \mid Y=y] {\rm d}y_1 + \frac{x_1^2}{2}, \label{eq:FinalIntegration}
\end{align}
where the labeled equalities follow from: a) using integration by parts;  b) using the identity in \eqref{eq:ScoreFunctionConditonalExpectation};  and  c) using the integral
\begin{align}
 \int_{\mathbb{R}}   y \left( Q(y_1)-    Q(y_1-x_1)  \right)  {\rm d} y= - \frac{x_1^2}{2}. 
 \end{align}

Next, observing that  $ \frac{Q(y_1)-    Q(y_1-x_1) }{-x_1}$ is a pdf since 
\begin{align}
\int_{\mathbb{R}}   \frac{Q(y_1)-    Q(y_1-x_1)}{- x_1} dy_1=1, 
\end{align}
and putting \eqref{Eq:diffofDenistities} and \eqref{eq:FinalIntegration} together we have that 
\begin{align}
i(x,P_X)-i(0,P_X)= -x_1 \E[  \E[X_1  \mid Y=W] ]  + \frac{x_1^2}{2},
\end{align} 
where $W$ is a random vector such that $W_1 \sim \frac{Q(y_1)-    Q(y_1-x_1) }{-x_1}$ and $W_i \sim \mathcal{N}(0,1)$  for $2 \le i \le n$. 

Combining \eqref{eq:FinalIntegration}   with   the conditional expectation of $X_R$ in  \eqref{eq:ConditionalExpectation}  and choosing $x_1=R$ the difference in \eqref{Eq:diffofDenistities} is given by 
\begin{align}
i(R,P_{X_R})-i(0,P_{X_R})=- R^2 \E \left[  \frac{W_1}{ \|W\| } \mathsf{h}_{\frac{n}{2}} \left(  R \|W\|  \right)   \right]  +\frac{R^2}{2}.  \label{Eq:ForX_R}
\end{align} 
The proof  is concluded by applying \eqref{Eq:ForX_R} to the sufficient and necessary condition in \eqref{eq:newCOndiitonForOptimality}.

\section{An Alternative Proof of the Lower Bound $\sqrt{n} \le \bar{R}_n$} 
\label{sec:AlternativeProof}
In this section we give an alternative proof of the lower bound  $\sqrt{n} \le \bar{R}_n$. The main idea is to show that  $x \mapsto i(x,P_X)$ is a subharmonic function where the notion of subharmonic functions is defined next. 
\begin{definition} \emph{(Subharmonic Function)}  Suppose that the function $f$ is twice continuously differentiable on an open set $G \in {{\mathbb  {R}}}^{n}$. Then $f$ is \emph{subharmonic} if   $ \nabla^2 f \geq 0$ on $G$ where  $\nabla^2$  is the Laplacian\footnote{ If $f$ is twice differentiable its Laplacian is given by by $\nabla^2 f= \sum_{i=1}^n \frac{\partial^2 f}{\partial x^2_i}$.}.
\end{definition}

We will use an important property that  a subharmonic function always attains its maximum on the boundary of a set  as shown in the following theorem. 
\begin{theorem} \label{thm:MaximumPrincipleOfSubharmonic} \emph{(Maximum Principle of  Subharmonic Functions)} Suppose that $G$ is a connected open set.  If $f$
is subharmonic and attains a global maximum value in  the interior of $G$, then $f$ is constant in $G$.
\end{theorem} 

To show the desired bound we will use  Theorem~\ref{thm:MaximumPrincipleOfSubharmonic} together with yet another identity that relates estimation and information measures \cite[Property 3]{hatsell1971some}. 

\begin{lem} Denote the likelihood function  by 
\begin{align}
\ell(y)\coloneqq \frac{f_Y(y)}{\frac{1}{(2\pi)^{\frac{n}{2}}} \eu^{-\frac{\|y\|^2}{2}}}.
\end{align}
Then,
\begin{align}
\nabla^2 \log\ell(y)&=   \E[  \|  X- \E[X \mid Y]\|^2  \mid Y=y  ]  \notag\\
&   \coloneqq  {\rm Var}(X  \mid Y=y). 
\end{align}
\end{lem}

The next result shows that the function $ x \mapsto i(x,P_X)$   is subharmonic if $X$ is contained in a small enough neighborhood. 
\begin{theorem}\label{thm:InformationDensityIsSubharmonic} Suppose that $X \in \mathcal{B}_0(R)$.   Then,  for $R \le \sqrt{n}$  and all $x  \in \mathcal{B}_0(R) $     the function    $ x \mapsto i(x,P_X)$   is subharmonic. 
\end{theorem} 
\begin{IEEEproof}
Observe that  $ i(x,P_X)$ can be written in terms of the log-likelihood function as follows: 
\begin{align}
& i(x,P_X) \notag\\%&= - \E \left[ \log f_Y(Y) | X=x \right]  - h(Z)\\
 &=  - \E \left[ \log f_Y(x+Z)  \right]  - h(Z)\\ 
  &=  - \E \left[  \log \ell(x+Z)   \right]    -  \E \left[ \log \frac{1}{(2\pi)^{\frac{n}{2}}} \eu^{-\frac{\|x+Z\|^2}{2}}  \right] - h(Z)\\
    &=  - \E \left[  \log\ell(x+Z)   \right]     +  \frac{\|x \|^2  }{2}. 
    \end{align}
       Therefore,  using the fact that $X \in   \mathcal{B}_0(R)$  we have that 
    \begin{align}
   \nabla^2  i(x,P_X) & =  -  \E \left[  {\rm Var}(X \mid Y) \mid X =x  \right]   +  n \\
   & \ge   -R^2+  n ,
    \end{align}
    where in the last step we have used the bound ${\rm Var}(X  \mid Y) \le R^2$. 
    This concludes the proof. 
\end{IEEEproof}

As a consequence of Theorem~\ref{thm:MaximumPrincipleOfSubharmonic} and Theorem~\ref{thm:InformationDensityIsSubharmonic} we have the following corollary.

\begin{corollary} For $R \le \sqrt{n}$ 
\begin{align}
\max_{X \in \mathcal{B}_0(R) } I(X;X+Z) =I(X_R;X_R+Z),
\end{align}
or,  alternatively,   $\sqrt{n} \le \bar{R}_n$. 
\end{corollary}

The  proof of Theorem~\ref{thm:InformationDensityIsSubharmonic} gives yet another example of the utility of identities between estimation and information measures.

\section{Discussion} 
\label{sec:Discussion}
In this work we have characterized conditions under which an input distribution uniformly distributed  over a single sphere achieves the capacity of a vector  Gaussian noise channel with a constraint that the input must lie in the $n$-ball of radius $R$. 
We have also  shown that the largest radius $\bar{R}_n$ for which it is still optimal to use a single sphere  grows as $\sqrt{n}$.   Moreover, the exact limit of $ \frac{\bar{R}_n}{\sqrt{n}}$ as $n \to \infty$ is found to be  $ \approx  1.861$. 

A number of methods that we have used throughout the paper relied on using estimation theoretic representations of  information measures such as the I-MMSE relationship. 
The  path  via estimation theoretic arguments allows us to contrast  optimization of the mutual information with that of  a similar problem of optimizing the MMSE, that is 
\begin{align}
\max_{X \in \mathcal{B}_0(R)} \mmse(X|Y). \label{eq:maxMMSE}
\end{align}
Distributions that maximize \eqref{eq:maxMMSE}  are referred to as \emph{least favorable prior distributions} and have been shown to have a spherical structure similar to that of  the distributions that maximize the mutual information; an interested reader is referred to  \cite{dytsoISIT2018leastFavorable} and \cite{marchand2004estimation} and references therein.  Moreover, the  conditions for the optimality of a single sphere distribution (i.e., the maximum radius $\bar{R}_n^{\text{MMSE}}$)  in \eqref{eq:maxMMSE} have been found in  \cite{dytsoISIT2018leastFavorable} and \cite{berry1990minimax} and are given by a solution to the following equation: 
\begin{align}
\E \left[    \mathsf{h}_{\frac{n}{2}}^2\left(  R   \|  Z \| \right) \right] + \E \left[    \mathsf{h}_{\frac{n}{2}}^2\left(   R   \| x+ Z \| \right) \right] =1, \label{eq:optimalMMSE}
\end{align}
where $\|x\|=R$.   It is pleasing to see the similarity between the optimality condition for the MMSE in \eqref{eq:optimalMMSE} and  the optimality condition for the mutual information in \eqref{eq:ExactR}.   Note, however, that  \eqref{eq:ExactR} could not have been derived directly from \eqref{eq:optimalMMSE} or  vice versa.
 The values of  $\bar{R}_n^{\text{MMSE}}$  are shown in Table~\ref{table:ValuesOfRn}. The code for the numerical computation of   $\bar{R}_n^{\text{MMSE}}$  and $\bar{R}_n$ can be found at  \cite{alex_dytso_2018_1401556}.

\begin{table}[!t]

\renewcommand{\arraystretch}{1.3}
\caption{Values of $\bar{R}_n$ and $\bar{R}_n^{\text{MMSE}}$.}
\label{table:ValuesOfRn}
\centering

\begin{tabular}{ |c|c|c| } 
\hline
Dimension $n$ &  $\bar{R}_n$  & $\bar{R}_n^{\text{MMSE}}$  \\
\hline
1	&	1.666	&	1.057	\\
\hline
2	&	2.454	&	1.535	\\
\hline
3	&	3.065	&	1.908	\\
\hline
4	&	3.580	&	2.223	\\
\hline
5	&	4.031	&	2.501	\\
\hline
6	&	4.438	&	2.751	\\
\hline
7	&	4.811	&	2.981	\\
\hline
8	&	5.158	&	3.195	\\
\hline
9	&	5.483	&	3.395	\\
\hline
10	&	5.789	&	3.585	\\
\hline
11	&	6.080	&	3.765	\\
\hline
12	&	6.359	&	3.936	\\
\hline
13	&	6.625	&	4.101	\\
\hline
14	&	6.881	&	4.259	\\
\hline
15	&	7.128	&	4.412	\\
\hline
16	&	7.367	&	4.560	\\
\hline
17	&	7.598	&	4.702	\\
\hline
18	&	7.823	&	4.841	\\
\hline
19	&	8.041	&	4.976	\\
\hline
20	&	8.254	&	5.107	\\
\hline
21	&	8.461	&	5.235	\\
\hline
22	&	8.663	&	5.360	\\
\hline
23	&	8.860	&	5.483	\\
\hline
24	&	9.054	&	5.602	\\
\hline
25	&	9.243	&	5.719	\\
\hline
26	&	9.428	&	5.834	\\
\hline
27	&	9.610	&	5.946	\\
\hline
28	&	9.789	&	6.056	\\
\hline
29	&	9.964	&	6.165	\\
\hline
30	&	10.136	&	6.271	\\
\hline
31	&	10.306	&	6.376	\\
\hline
32	&	10.472	&	6.479	\\
\hline
33	&	10.636	&	6.580	\\
\hline
34	&	10.798	&	6.680	\\
\hline
35	&	10.957	&	6.779	\\
\hline
\end{tabular}
\end{table}

\begin{figure}
\center

\pgfplotsset{every axis plot/.append style={very thick}}
\begin{tikzpicture}

\begin{axis}[%
width=6.953cm,
height=6.226cm,
at={(1.159in,0.77in)},
scale only axis,
xmin=0,
xmax=35,
xlabel style={font=\color{white!15!black}},
xlabel={$n$},
ymin=1,
ymax=11,
axis background/.style={fill=white},
xmajorgrids,
ymajorgrids,
legend style={at={(0.1,0.712)}, anchor=south west, legend cell align=left, align=left, draw=white!15!black}
]
\addplot [color=black, dashed]
  table[row sep=crcr]{%
1	1.66592407226562\\
2	2.45290612235946\\
3	3.06515589156213\\
4	3.57968139648438\\
5	4.03161664259876\\
6	4.43727070272028\\
7	4.81080558750934\\
8	5.15746871695396\\
9	5.48295593261719\\
10	5.78938653480014\\
11	6.08053782460932\\
12	6.35815617417729\\
13	6.62498590491425\\
14	6.88123884623733\\
15	7.12855040961604\\
16	7.36770629882812\\
17	7.59842137888772\\
18	7.82278956101171\\
19	8.04094340864291\\
20	8.25338231532763\\
21	8.46062652701911\\
22	8.66280484559857\\
23	8.86027971288904\\
24	9.05420902619188\\
25	9.24312591552734\\
26	9.42835441814296\\
27	9.61033686950775\\
28	9.78870658329041\\
29	9.96402575577891\\
30	10.1361184136208\\
31	10.3052831013351\\
32	10.4726812867666\\
33	10.636022327417\\
34	10.797928899937\\
35	10.9572864091877\\
};
\addlegendentry{$\bar{R}_n$}

\addplot [color=black, dashdotted]
  table[row sep=crcr]{%
1	1.05673217773438\\
2	1.53510016479641\\
3	1.90798864671051\\
4	2.22323608398438\\
5	2.50087088847049\\
6	2.75137769219205\\
7	2.98135511124892\\
8	3.19505750232064\\
9	3.39537048339844\\
10	3.58492155093704\\
11	3.76483169123099\\
12	3.93646709493021\\
13	4.1011660317051\\
14	4.25940671556971\\
15	4.41209605640909\\
16	4.55953979492188\\
17	4.70238258862844\\
18	4.84130126686379\\
19	4.97609258896587\\
20	5.10734221624704\\
21	5.23535638648813\\
22	5.36034735039697\\
23	5.48250274369548\\
24	5.60198932095467\\
25	5.71895599365234\\
26	5.83353632447646\\
27	5.94600916614151\\
28	6.05633048144496\\
29	6.16468088902662\\
30	6.27132132092192\\
31	6.37592107824476\\
32	6.47897809439122\\
33	6.58048521853983\\
34	6.68033516913944\\
35	6.77867591907692\\
};
\addlegendentry{$\bar{R}_n^{\text{MMSE}}$}

\addplot [color=black,dotted]
  table[row sep=crcr]{%
1	1\\
2	1.4142135623731\\
3	1.73205080756888\\
4	2\\
5	2.23606797749979\\
6	2.44948974278318\\
7	2.64575131106459\\
8	2.82842712474619\\
9	3\\
10	3.16227766016838\\
11	3.3166247903554\\
12	3.46410161513775\\
13	3.60555127546399\\
14	3.74165738677394\\
15	3.87298334620742\\
16	4\\
17	4.12310562561766\\
18	4.24264068711928\\
19	4.35889894354067\\
20	4.47213595499958\\
21	4.58257569495584\\
22	4.69041575982343\\
23	4.79583152331272\\
24	4.89897948556636\\
25	5\\
26	5.09901951359278\\
27	5.19615242270663\\
28	5.29150262212918\\
29	5.3851648071345\\
30	5.47722557505166\\
31	5.56776436283002\\
32	5.65685424949238\\
33	5.74456264653803\\
34	5.8309518948453\\
35	5.91607978309962\\
};
\addlegendentry{$\sqrt{n}$}

\node[below right, align=left, draw=black]
at (rel axis cs:0.43,0.32) {LFD's are optimal};
\node[below right, align=left, draw=black]
at (rel axis cs:0.5,0.68) {LFD's are not optimal};

\end{axis}

%\begin{axis}[%
%width=8.917in,
%height=7in,
%at={(0in,0in)},
%scale only axis,
%xmin=0,
%xmax=1,
%ymin=0,
%ymax=1,
%axis line style={draw=none},
%ticks=none,
%axis x line*=bottom,
%axis y line*=left
%]
%\node[below right, align=left, draw=black]
%at (rel axis cs:0.431,0.257) {LFD's are optimal};
%\node[below right, align=left, draw=black]
%at (rel axis cs:0.446,0.513) {LFD's are not optimal};
%\end{axis}
\end{tikzpicture}%
\caption{Comparison of  $\bar{R}_n$, $\bar{R}_n^{\text{MMSE}}$ and $\sqrt{n}$.   For   $R  \le \bar{R}_n^{\text{MMSE}}$ the least favorable distributions (LPF's) are capacity achieving (optimal for short) and not capacity achieving for $R >\bar{R}_n^{\text{MMSE}}$.   }
\label{fig:compRMaxRmaxMMSE}
\end{figure}
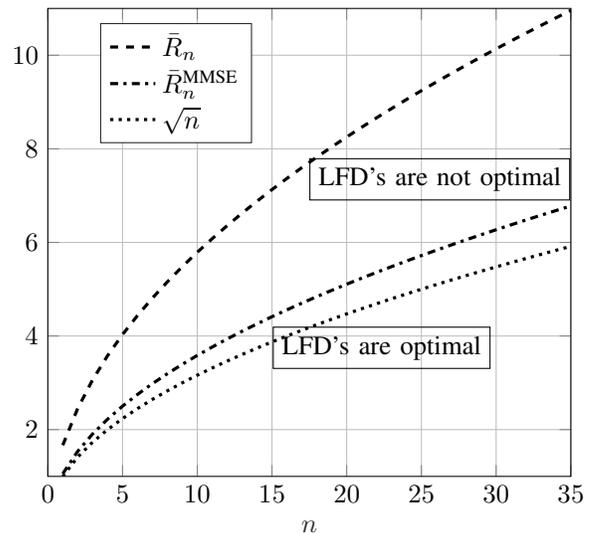

It is also interesting to point out that that $\bar{R}_n^{\text{MMSE}}$ is always lagging behind   $\bar{R}_n$  as we increase $n$ as shown in Fig.~\ref{fig:compRMaxRmaxMMSE}. Notably this behavior persists even as $n \to \infty$  since for the MMSE\footnote{ The exact limit is given by  $\lim_{n \to \infty} \frac{\bar{R}_n^{\text{MMSE}}}{\sqrt{n}}= c_{\text{MMSE}}=\frac{\sqrt{\sqrt[3]{9-\sqrt{69}}+\sqrt[3]{9+\sqrt{69}}}}{\sqrt[6]{2} \sqrt[3]{3}}  \approx    1.15096$; see   \cite{dytsoISIT2018leastFavorable} and \cite{marchand2002minimax}  for the details.   }
\begin{align}
\lim_{n \to \infty} \frac{\bar{R}_n^{\text{MMSE}}}{\sqrt{n}}= c_{\text{MMSE}}  \approx    1.151, 
\end{align}
where   $c_{\text{MMSE}}$ is the solution of the following equation:
\begin{align}
 \frac{   c^2 }{ \left( \frac{1}{2 }  +\sqrt{ \frac{1}{4 }  + c^2} \right)^2} \hspace{-0.05cm}  + \hspace{-0.05cm}\frac{   c^2(1+ c^2) }{ \left( \frac{1}{2 }  +\sqrt{ \frac{1}{4 }  +  c^2(1+  c^2)} \right)^2}=1, \label{eq:OptimalcMMSE}
\end{align}
while  for the mutual information according to Theorem~\ref{thm:MainResult}
\begin{align}
\lim_{n \to \infty} \frac{\bar{R}_n}{\sqrt{n}}  \approx    1.861.
\end{align}
Again, note the similarity between \eqref{eq:OptimalcMMSE} and \eqref{eq:SolutionInfinity}. 

The lagging of $\bar{R}_n^{\text{MMSE}}$ behind $\bar{R}_n$ also points out that the following bounding technique,  which relies on the I-MMSE, results in a tight bound if $R \le \bar{R}_n^{\text{MMSE}}$ and is not  tight if    $\bar{R}_n^{\text{MMSE}} \le R \le \bar{R}_n$:
\begin{align}
\max_{X  \in \mathcal{B}_0(R)} I(X;Y)  &= \max_{X  \in \mathcal{B}_0(R)} \frac{1}{2} \int_0^1 \mmse(X|Y_\gamma) {\rm d} \gamma\\
& \le   \frac{1}{2} \int_0^1  \max_{X  \in \mathcal{B}_0(R)}  \mmse(X|Y_\gamma) {\rm d} \gamma, \label{eq:anflanflkajflaj}
\end{align}
Such a condition for tightness of the bound via the I-MMSE relation was already pointed out in   \cite{raginsky2008information} for $n=1$.   Interestingly for several multiuser problems \cite{BustinMMSEbadCodes,MMSEdisturbance,dytso2017view}, with a second moment constraint on the input instead of an amplitude constraint, such lagging vanishes as $n \to \infty$ and  bounds via the I-MMSE of the type in \eqref{eq:anflanflkajflaj} (i.e., where the maximum and the integral are interchanged)  are tight.   The fundamental difference is that in the aforementioned works the Gaussian distribution is optimal in the limit of $n$, while in \eqref{eq:Capacity} and \eqref{eq:maxMMSE} Gaussian inputs are not optimal even as $n \to \infty$.

The optimality of an input distribution on a single sphere also has important practical implications as it suggests that phase modulation is optimal.  Note  that in practice, however, the continuous sphere would have to be discretized.  The accuracy of such a discretization can potentially be evaluated  by using the fact that  the mutual information is continuous in  the Wasserstein metric  over a set of distributions with compact support \cite{FunctionalPropMMSE}.  

 An  ambitious future direction is to consider an extension of the result in this paper to a general MIMO channel.  For a recent survey on discrete inputs in MIMO systems the interested reader is referred to \cite[Corollary 4]{wu2018survey}. 
 
 Another interesting direction is to consider an input average power constraint (i.e., $\E[\|X\|^2] \le P$) together with  the input amplitude constraint analyzed in this paper.

%\appendix

\begin{appendices}

\section{On the Numerical Evaluation of the Integrals in \eqref{eq:SecondCondition} and \eqref{eq:optimalMMSE}}
\label{app:sec:ComputationOfRatioOfBesselFunctions}

Evaluation of the expectations  in \eqref{eq:SecondCondition} and \eqref{eq:optimalMMSE} for any given $R$ may be performed using Monte-Carlo methods. The ratios of Bessel functions in the expectations can be evaluated precisely thanks to the known continued fraction expansion \cite{cuyt2008handbook}
\begin{equation}
\mathsf{h}_n(x)=\frac{\mathsf{I}_n(x)}{\mathsf{I}_{n-1}(x)}=\frac{1}{b_1+\frac{1}{b_2+\cdots}},
\end{equation}
where
\begin{equation}
b_k=\frac{2(n+k-1)}{x}. 
\end{equation}

\begin{figure}

\begin{subfigure}[a]{0.45\textwidth}
\center
            \input{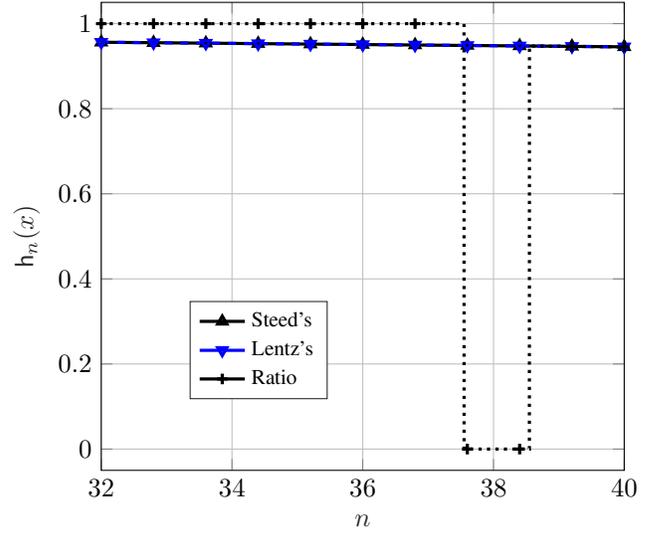}
\caption{Plot of $\mathsf{h}_n(x)$ vs. $n$ for $x=705$. }
\label{fig:RatioBessel:DenominatorOverflow}
        \end{subfigure}%

\begin{subfigure}[c]{0.45\textwidth}
\center
            \input{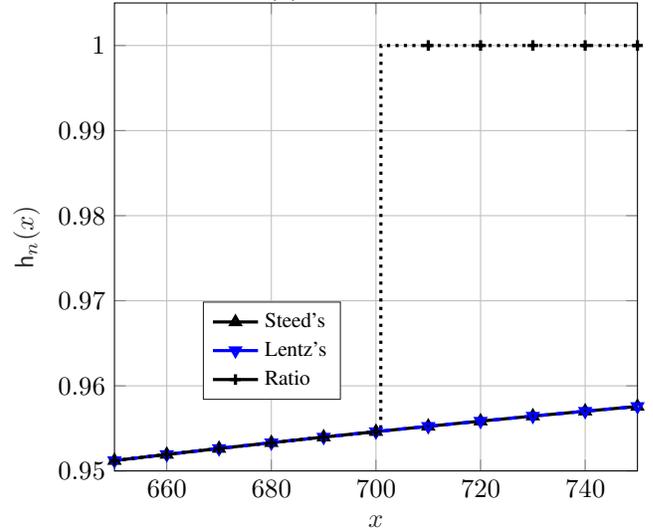}
\caption{Plot of $\mathsf{h}_n(x)$ vs. $x$ for   $n=33$.}
        \end{subfigure}%
        
        \caption{Comparison of values of $\mathsf{h}_n(x)$  obtained via Steed's algorithm, Lentz's algorithm and direct evaluation of the ratio of Bessel functions.}
\label{fig:RatioExample}
\end{figure}

The continued fraction can be evaluated via Steed's or Lentz's algorithm \cite{press2007numerical}, either of which gives  stable and accurate results, whereas the direct evaluation of the ratio of Bessel functions may lead to  floating-point overflows at high values of $x$. An example of the overflow for double precision  values is shown on Fig.~\ref{fig:RatioExample}.   Note that in Fig.~\ref{fig:RatioBessel:DenominatorOverflow} the zero values of the function $\mathsf{h}_n(x)=\frac{\mathsf{I}_n(x)}{\mathsf{I}_{n-1}(x)}$ around $n=38$ correspond to denominator overflows while the one values  for smaller values of $n$ correspond to both numerator and denominator overflows.  Moreover, observe that neither Steed's or Lentz's algorithm experiences this issue.

Since $h_n(x)$ is a monotonically increasing function of $x$, the integrands in  \eqref{eq:SecondCondition}   and \eqref{eq:optimalMMSE} are monotonically increasing functions of $R$. Hence, given lower and upper bounds on $R$, the zeroes of  \eqref{eq:SecondCondition}   and \eqref{eq:optimalMMSE} may be obtained via binary searches. In our simulations, we set the lower and upper bounds to $\sqrt{n}$ and $3\sqrt{n}$, respectively. Note that this interval includes both the set $[1.6659\sqrt{n}, 1.8609\sqrt{n}]$  for the maximum mutual information setting, and the set $[1.0567\sqrt{n}, 1.151\sqrt{n}]$ for the maximum MMSE setting.

We sampled $10^8$ chi-square samples while evaluating the expectations for the binary searches. During the evaluation of \eqref{eq:SecondCondition}, we integrated directly over the distribution of $W_1$, and we distributed the chi-square samples uniformly across the effective domain of $W_1$, which was set as $[-7,R+7]$ to capture all but a negligible amount of the probability mass. During the evaluation of \eqref{eq:optimalMMSE}, we sampled evenly from the central and non-central chi-square distributions\footnote{Alternatively, more samples can be taken from the non-central chi-square due to its higher variance.}.

The $\bar{R}_n$ and $\bar{R}_n^{\text{MMSE}}$ values reported in Table~\ref{table:ValuesOfRn} have consistently led to residuals well below $10^{-4}$ during the Monte-Carlo evaluations of the integral equations. In our experience, multiple binary searches in this setting do not lead to changes in $\bar{R}_n$ and $\bar{R}_n^{\text{MMSE}}$ values beyond the fourth digit after the decimal point. Interested readers may refer to our implementation and simulation data
found at \cite{alex_dytso_2018_1401556}. 

%
%Since $\mathsf{h}_n(x)$ is a monotonically increasing function of $x$, the integrand  in \eqref{eq:SecondCondition}  is a monotonically increasing function of $R$. Hence, given a lower and upper bound on $R$, the zero of \eqref{eq:SecondCondition} can be obtained via binary search. In our simulations, we set the lower and upper bounds to $ \sqrt{n} $ and $ 3 \sqrt{n}$, respectively.  We sampled $10^8$ chi-square samples while evaluating the expectations for the binary searches. The $\bar{R}_n$ and $ \bar{R}_n^{\text{MMSE}}$  values reported in Table~\ref{table:ValuesOfRn}  have consistently led to residuals well below $10^{-4}$ during the Monte Carlo evaluations of the integral equations. In our experience, multiple binary searches in this setting do not lead to changes in $\bar{R}_n$ and $\bar{R}_n^{\text{MMSE}}$ values beyond the fourth digit after the decimal point. Interested readers may refer to our implementation  and simulation data found at \cite{alex_dytso_2018_1401556}. 

\section{Converting to Spherical Coordinates in \eqref{eq:aklfnlkandf;al}}
\label{sec:SphericalCoordinates}
Using that 
\begin{align*}
  \rho(y)= \left(- \|y\|  +     \frac{   R    \mathsf{I}_{ \frac{n}{2}}(\|y\| R)  }{   \mathsf{I}_{ \frac{n}{2}-1}(\|y\| R) } \right) \frac{y_1}{\|y\|}= M(\|y\|)   \frac{y_1}{\|y\|},
\end{align*}
we have
\begin{align*}
-g^\prime(x_1)& = \int_{\mathbb{R}^n}  \frac{1}{ (2 \pi)^{\frac{n}{2}}}  \eu^{- \frac{\sum_{i=2}^n y_i^2 + (y_1-x_1)^2}{2}}   M(\|y\|)   \frac{y_1}{\|y\|}
 {\rm d}y.
\end{align*}

Transforming $y_1,y_2, \dots ,y_n$ to the spherical coordinates $r, \phi_1, \dots,\phi_{n-1}$ where $r\ge 0$,  $0 < \phi_1 \le 2 \pi$ and $0<  \phi_i\le \pi$ for $i=2,\dots, n-1$ we have that 
\begin{subequations}
\begin{align}
y_1&= r \cos(\phi_1),\\
y_i&= r \cos(\phi_i)  \prod_{k=1}^{i-1} \sin \phi_k,  \, i=2, \dots , n-1,\\
y_n&=r  \prod_{k=1}^{n-1} \sin \phi_k,
\end{align}
and the Jacobian is given by 
\begin{align}
{\rm d}y=   r^{n-1}  \prod_{k=1}^{n-1} \left(\sin\phi_k \right)^{n-1-k} {\rm d}r {\rm d}\phi_1 \dots {\rm d}\phi_{n-1}.
\end{align}
\label{eq:SphericalCoordinates}
\end{subequations} 
Therefore, the derivative can be written as follows: 
\begin{align*}   % the steps are here. I commented them out 
&-g^\prime(x_1) \notag\\
%& = \int_0^{2\pi} \int_0^\pi ... \int_0^\pi  \int_{0}^r  \frac{1}{ (2 \pi)^{\frac{n}{2}}}  \eu^{- \frac{\|y -x\|^2}{2}}   M(r)   \frac{r \cos(\phi_1)}{r} 
%r^{n-1}  \prod_{k=1}^{n-1}   \left(\sin\phi_k \right)^{n-1-k} {\rm d}r {\rm d}\phi_1...{\rm d}\phi_{n-1}\\
& \stackrel{a)}{=}  \prod_{k=2}^{n-1} \int_0^\pi  \left(\sin\phi_k \right)^{n-1-k} {\rm d} \phi_k  \int_0^{2\pi}  \int_{0}^\infty \frac{\eu^{- \frac{r^2-2rx_1\cos(\phi_1) + x_1^2}{2}}}{ (2 \pi)^{\frac{n}{2}}}  \notag\\
& \cdot        M(r) \cos(\phi_1) 
r^{n-1}  \left(\sin\phi_1 \right)^{n-1}  {\rm d}r {\rm d}\phi_1\\
%& =  \prod_{k=2}^{n-1} \int_0^\pi  \left(\sin\phi_k \right)^{n-1-k} {\rm d} \phi_k    \int_{0}^r  \frac{1}{ (2 \pi)^{\frac{n}{2}}}  \eu^{- \frac{r^2+ x_1^2}{2}}   M(r)  \int_0^{2\pi}   \eu^{ r x_1\cos(\phi_1) } \cos(\phi_1)  \left(\sin\phi_1 \right)^{n-1}
%{\rm d}\phi_1 r^{n-1}  {\rm d}r \\
& \stackrel{b)}{=} S_{n-2}   \int_{0}^\infty  \frac{ \eu^{- \frac{r^2+ x_1^2}{2}}}{ (2 \pi)^{\frac{n}{2}}}   \frac{  M(r)  2^{\frac{n}{2}} \sqrt{\pi} \Gamma \left( \frac{n-1}{2}\right)}{ (x_1 r)^{\frac{n-2}{2}}} \mathsf{I}_{\frac{n}{2}}(x_1 r) r^{n-1}  {\rm d}r \\
%& = S_{n-2} 2^{\frac{n-2}{2}} \sqrt{\pi} (n-3) \Gamma \left( \frac{n-3}{2}\right)   \int_{0}^r  \frac{1}{ (2 \pi)^{\frac{n}{2}}}  \eu^{- \frac{r^2+ x_1^2}{2}}   M(r) \left( \frac{r}{x } \right)^{\frac{n}{2}}  x\mathsf{I}_{\frac{n}{2}}(x_1 r )   {\rm d}r \\
& \stackrel{c)}{=}       x \int_{0}^\infty  \eu^{- \frac{r^2+ x_1^2}{2}}   M(r)  \left( \frac{r}{x } \right)^{\frac{n}{2}}  \mathsf{I}_{\frac{n}{2}}(x_1 r )   {\rm d}r \\
& =      2 x \int_{0}^r     M(r)   \eu^{- \frac{r^2+ x_1^2}{2}} \frac{1}{2}\left( \frac{r}{x } \right)^{\frac{n}{2}}  \mathsf{I}_{\frac{n}{2}}(x_1 r )   {\rm d}r \\
& \stackrel{d)}{=}       2 x  \E \left[ M \left( \sqrt{V^2} \right) \right],
\end{align*}
where the labeled equalities follow from: a) using spherical coordinates in \eqref{eq:SphericalCoordinates}; b) using that
  \begin{align*}
   \prod_{k=2}^{n-1} \int_0^\pi  \left(\sin\phi_k \right)^{n-1-k} {\rm d} \phi_k =S_{n-1},
   \end{align*}
   and that 
   \begin{align*}
    &\int_0^{2\pi}   \eu^{ r x_1\cos(\phi_1) } \cos(\phi_1)  \left(\sin\phi_1 \right)^{n-1}  {\rm d}\phi_1 \notag\\
    &= \frac{2^{\frac{n}{2}} \sqrt{\pi} \Gamma \left( \frac{n-1}{2}\right)}{ (x_1 r)^{\frac{n-2}{2}}} \mathsf{I}_{\frac{n}{2}}(x_1 r) r^{n-1};
   \end{align*}
   c) using that  $ S_{n-2} 2^{\frac{n}{2}} \sqrt{\pi}  \Gamma \left( \frac{n-1}{2}\right)=(2 \pi)^{\frac{n}{2}}$; and observing that  $\frac{1}{2} \eu^{- \frac{r^2+ x_1^2}{2}}\left( \frac{r}{x } \right)^{\frac{n}{2}}   \mathsf{I}_{\frac{n}{2}}(x_1 r )$ is the pdf of a chi-square random variable of degree $n+2$ with non-centrality $x_1^2$. 

This concludes the proof. 
\end{appendices}

\bibliography{refs}
\bibliographystyle{IEEEtran}
\end{document}